\documentclass[twocolumn]{aastex631}
\renewcommand{\edit}[2]{#2}
\usepackage{amsmath}
\makeatletter
\AtBeginDocument{%
  \ifnumlines\else

  \fi
}
\makeatother

\begin{document}

\title{Revisiting the greenhouse effect of non-greenhouse gases in the atmospheres of Earth-like planets}

\author[0000-0002-6602-7113]{Tetsuo Taki}
\correspondingauthor{Tetsuo Taki}
\email{takitetsuo@gmail.com}
\affiliation{Department of General Systems Studies, Graduate School of Arts and Sciences, The University of Tokyo, Tokyo 153-8902, Japan}

\author[0000-0003-1965-1586]{Hiroyuki Kurokawa}
\affiliation{Department of General Systems Studies, Graduate School of Arts and Sciences, The University of Tokyo, Tokyo 153-8902, Japan}
\affiliation{Department of Earth and Planetary Science, Graduate School of Science, The University of Tokyo, Tokyo 113-0033, Japan}

\author[0000-0002-2786-0786]{Yuka Fujii}
\affiliation{Division of Science, National Astronomical Observatory of Japan, Tokyo, 181-8588, Japan}

\author{Kosuke Aoki}
\affiliation{Department of Earth and Planetary Sciences, Institute of Science Tokyo, Tokyo 152-8551, Japan}

%%%%%
% < 250 words
\begin{abstract}
% 239 words
Although non-greenhouse gases can vary substantially in abundance in Earth-like atmospheres, their climatic influences remain insufficiently understood.
To investigate how such gases regulate climate, we vary the abundance of N$_2$ as a representative non-greenhouse component in one-dimensional N$_2$--CO$_2$--H$_2$O model atmospheres.
Beyond pressure broadening of absorption lines and Rayleigh scattering emphasized in previous studies, our results show that changes in background N$_2$ pressure influence climate by modifying the amount of atmospheric H$_2$O, producing two effects: altering the thermodynamic lapse rate (H$_2$O-dilute warming) and changing the radiative contribution of H$_2$O to the greenhouse effect (H$_2$O-load warming).
The resulting climate response to increasing N$_2$ depends on the CO$_2$ abundance.
Under low CO$_2$ conditions, dilution of atmospheric H$_2$O leads to warming, whereas under high CO$_2$ conditions, increased H$_2$O loading also produces warming.
At sufficiently high N$_2$ abundances, Rayleigh scattering induces cooling, an effect further amplified by the accompanying decrease in atmospheric H$_2$O.
Under high CO$_2$ conditions, however, enhanced H$_2$O loading increases the absorption of stellar radiation and overwhelms the contribution of Rayleigh scattering, causing the cooling response to disappear.
These results reveal multiple physical pathways through which non-greenhouse gases influence climate and provide a framework for understanding climate responses and habitability in diverse Earth-like atmospheres.
\end{abstract}

\keywords{Exoplanet atmospheres(487);  Planetary atmospheres (1244); Atmospheric composition (2120); Exoplanet atmospheric composition (2021); Planetary science (1255)}

%%%%%%
\section{Introduction} \label{sec:intro}

Atmospheric pressure and composition of terrestrial planets largely control their climates \citep[][and references therein]{catling2017atmospheric}. 
Greenhouse gases contribute to the warming of planetary surfaces. 
For instance, CO$_2$ and H$_2$O keep the current Earth warm enough to sustain surface water; otherwise, Earth's oceans could be fully covered by ice. 
Atmospheric compositions influence the structures of planetary atmospheres as well.
The temperature lapse rate in the troposphere on current Earth is characterized by the moist rather than dry adiabatic lapse rates ($\simeq$$-6\ \mathrm{K\ km^{-1}}$ vs. $\simeq$$-10\ \mathrm{K\ km^{-1}}$), where the latent-heat release in water vapor condensation makes significant difference. 
Moreover, the vertical profile of water vapor plays an important role in atmospheric chemistry, through the photolysis to produce OH radicals, and in the retention of surface water against atmospheric escape.

Non-greenhouse gases such as N$_2$, CO, and O$_2$ are also known to influence planetary climates.
\cite{2009NatGe...2..891G} proposed that a higher N$_2$ level helped warming early Earth under the faint young sun; their climate model showed that a doubling of present N$_2$ level causes warming by $\simeq +4\ \mathrm{^\circ C}$. 
While a low $p_{{\rm N}_2}$, the partial pressure of N$_2$, suggested later from geological and geochemical record \citep{Som+2012Natur.484..359S,Som+2016NatGe...9..448S,Marty+2013Sci...342..101M} may not support its role to solve the faint young sun paradox in the Archean \citep[see also][for the case against the constraint from fossil raindrop imprints]{Kavanagh+Goldblatt2015E&PSL.413...51K}, a large budget of nitrogen in the crust and mantle \citep{2009NatGe...2..891G,Marty2012E&PSL.313...56M,Johnson+Goldblatt2015ESRv..148..150J,hirschmann2016constraints} indicates that Earth started with a dense, N$_2$-rich atmosphere \citep{Sakuraba+2021NatSR..1120894S,Kurokawa+2022GGG....2310295K}, which could have influenced Hadean Earth's climate. 
The atmospheric inventory of N$_2$ and other noncondensible gases is important for the cold trap to operate as well \citep{2009NatGe...2..891G,Wordsworth+Pierrehumbert2014ApJ...785L..20W}. 
As a consequence, the climate effects of N$_2$ and other non-greenhouse gases influence the extent of the habitable zone \edit1{\citep[e.g.,][]{Vladilo+2013ApJ...767...65V,2020MNRAS.494..259R}}.

Atmospheres with rich non-greenhouse gases, like the current Earth's, might be common in the early solar system as well as extrasolar terrestrial worlds, amplifying the importance of studying their climate effects. 
Warming by N$_2$ has been suggested to be important on early Mars \citep{vonParis2013P&SS...82..149V,Hu+Thomas2022NatGe..15..106H,Thomas+2023PSJ.....4...41T,adams2025episodic}. 
N$_2$-rich atmospheres have also been suggested for early, potentially-habitable Venus \citep{Way+2016GeoRL..43.8376W,Way+DelGenio2020JGRE..12506276W} and hypothetical extrasolar Earth-like planets with active carbonate-silicate cycling \citep{Lammer+2019AsBio..19..927L}. 
Other non-greenhouse gases potentially rich on some terrestrial planets include CO and O$_2$. 
On modern Earth, atmospheric O$_2$ originates from photosynthesis. 
Water photolysis and hydrogen escape can also leave O$_2$-rich atmospheres \citep{Wordsworth+Pierrehumbert2014ApJ...785L..20W,Luger+Barnes2015AsBio..15..119L}. 
Runaway photodissociation of CO$_2$ can produce atmospheres rich in CO and O$_2$ on planets orbiting M-type stars \citep{Tian+2014,harman2015abiotic,Hu+2020}, in colder environments such as early Mars \citep{zahnle2008photochemical}, or with volcanic supply of reducing gases \citep{Sholes+2017Icar..290...46S,Watanabe+Ozaki2024ApJ...961....1W}.

Despite its potential importance, the climate effects of non-greenhouse gases and their mechanisms have not been studied systematically. 
\edit1{Previous studies have shown that increasing the abundance of non-greenhouse gases such as N$_2$ can produce either warming or cooling at the planetary surface, depending on the atmospheric state.}
\edit1{For example, increasing N$_2$ can warm} early Earth and Mars \edit1{through} pressure broadening of absorption lines of conventional greenhouse gases \citep{2009NatGe...2..891G,vonParis2013P&SS...82..149V,Hu+Thomas2022NatGe..15..106H}.
\edit1{By contrast,} \cite{Keles+2018AsBio..18..116K} showed that increasing the atmospheric pressure beyond $4\ \mathrm{bar}$ causes cooling of the surface for Earth-like planets due to Rayleigh scattering. 
\cite{2013ApJ...778..154W} \edit1{also} showed that increasing N$_2$ causes either warming \edit1{or} cooling depending on $p_{\mathrm{CO}_2}$.
\edit1{
Related studies near the runaway greenhouse limit further suggest that non-greenhouse gases can affect climate in additional ways, for example by strengthening H$_2$O greenhouse warming in H$_2$-rich atmospheres \citep{2019ApJ...881..120K} or producing an overshoot in outgoing longwave radiation (OLR) compared to the Simpson--Nakajima threshold in N$_2$+H$_2$O atmospheres \citep{2022A&A...658A..40C}.
}
\edit1{These studies have identified} several relevant mechanisms, including pressure broadening, Rayleigh scattering, and water vapor feedback \citep{2009NatGe...2..891G}, \edit1{but the} limited parameter spaces studied hinder us from understanding when warming or cooling dominates and what is the dominant mechanism.

This study aims to elucidate the effects of non-greenhouse gases on planetary climates: the surface temperature and vertical profiles of temperature and water vapor. 
For this purpose, we perform an extensive parameter study in $p_{\mathrm{CO}_2}$ and $p_{{\mathrm N}_2}$ and analyze the influences of N$_2$ on the climate. 
We show that, in addition to i) pressure broadening and ii) Rayleigh scattering emphasized by previous studies, the climate effects of non-greenhouse gases are largely determined by water vapor feedback, which can further be classified into iii) the change in the adiabatic lapse rate (H$_2$O-dilute warming), and iv) the greenhouse effect of H$_2$O vapor (H$_2$O-load warming); see Section~\ref{sec:origin-warming-trend} for the terminology and the rationale behind it for the latter two effects.
We use N$_2$ as a proxy for non-greenhouse gases, but our conclusions are applicable quantitatively to CO, which has the same molecular weight, and at least qualitatively to other non-greenhouse gases. 
Section~\ref{sec:method} presents our methods. 
Section~\ref{sec:results} shows our results. 
Section~\ref{sec:discussions} discusses model limitations, comparison with previous studies, and implications for different planetary environments. 
We conclude in Section~\ref{sec:conclusion}.

\section{Methods} 
\label{sec:method}

\subsection{Model atmosphere}
\label{sec:climate-model}

We employ a one–dimensional climate model that computes the atmospheric structure in radiative balance at the top of the atmosphere (TOA).
The model consists of an adiabatic troposphere and an isothermal stratosphere.
We describe the compositions of our model atmospheres and the roles of the individual species in Section~\ref{sec:atmosph-comp}.
We then introduce the physical structure of the model atmospheres in Section~\ref{sec:atmosph-struct}.
In that section, we place particular emphasis on the temperature structure and the factors that determine it.

\subsubsection{Atmospheric composition}
\label{sec:atmosph-comp}

In this study, we consider two types of model atmospheres, namely the N$_2$–CO$_2$–H$_2$O atmosphere as the fiducial case, which mimics the atmospheres of Earth-like planets, and the N$_2$–CO$_2$ atmosphere as the dry-atmosphere case.
We test the climate response to an increase in N$_2$ as a non-greenhouse gas.
By comparing the responses of the two model atmospheres, we identify the contributions of four effects associated with an increase in N$_2$, namely i) pressure broadening, ii) Rayleigh scattering, iii) the change in the adiabatic lapse rate (H$_2$O-dilute warming), and iv) the greenhouse effect of H$_2$O vapor (H$_2$O-load warming).
As mentioned in Section~\ref{sec:intro}, the latter two effects originate from H$_2$O, and therefore they are absent in the dry-atmosphere case.
Unless otherwise stated, descriptions and discussions of H$_2$O-related quantities, such as relative humidity (RH), are based on the fiducial case.

Since N$_2$ is treated as a noncondensible and non-greenhouse gas, it makes no direct contribution to the absorption of atmospheric radiation. 
We ignored collision induced absorption (CIA) of N$_2$ for our purpose \edit1{(see also Section~\ref{subsec:discussion:limitations})}.
N$_2$ directly contributes to the cooling of the surface temperature through the Rayleigh scattering.
In addition, an increase in N$_2$ broadens the absorption lines of the greenhouse gases through pressure broadening, and this effect results in indirect warming.

Both CO$_2$ and H$_2$O, which are greenhouse gases, contribute to the absorption and scattering of radiation.
The main difference between these two species in our model lies in how their atmospheric abundances are determined.
Because H$_2$O is a condensible gas under the typical pressure and temperature conditions of Earth-like planetary atmospheres, its abundance at each altitude depends on the local atmospheric temperature.
In contrast, we treat CO$_2$ as a noncondensible gas because CO$_2$ condensation has only a limited impact on our results compared to the condensation of H$_2$O.

\subsubsection{Atmospheric structure}
\label{sec:atmosph-struct}

We assume that the planetary atmosphere consists of two parts: a troposphere, where the temperature follows the pseudo-moist adiabatic lapse rate, and a perfectly isothermal stratosphere at $200$ K
\edit1{, following the standard treatment adopted by \citet{1988Icar...74..472K} and subsequent studies.}

Since we focus on the atmospheric structure in terms of the temperature–pressure relation, we define the inverse of the pseudo-moist adiabatic lapse rate, $\Gamma^{-1}_{\rm moist}$, following \citet{1988Icar...74..472K}, for reference
\footnote{\edit1{We use the inverse lapse rate, ${\rm d}\ln p/{\rm d}\ln T$, rather than the conventional lapse rate because it directly represents the slope of the atmospheric profile in the $T$--$p$ plane and facilitates comparison with previous 1D studies following \citet{1988Icar...74..472K}. Readers accustomed to the conventional lapse rate should note that qualitative descriptions such as ``steeper'' and ``shallower'' are reversed relative to that notation.}}
:
\begin{eqnarray}
 \Gamma^{-1}_{\rm moist} &=&
  \frac{{\rm d}\ln p}{{\rm d}\ln T} \nonumber \\
 &=&
  \frac{p_{\rm v}}{p}\frac{{\rm d}\ln p_{\rm v}}{{\rm d}\ln T}+
  \frac{p_{\rm n}}{p}\left(
      1+\frac{{\rm d}\ln \rho_{\rm v}}{{\rm d}\ln T}
       -\frac{{\rm d}\ln \alpha_{\rm v}}{{\rm d}\ln T}
                     \right),
\label{eq:adiabatic-lapse-rate}
\end{eqnarray}
where $T$ is the temperature, $p$ is the pressure, and $p_i$ is the pressure that component $i$ would have if it existed alone \citep[see also][]{2025PSJ.....6..212L}, with $\rho_i$ being the density of component $i$.
Here we decompose the atmosphere into two components, referred to as a condensible gas (H$_2$O, $i=\mathrm{v}$) and a noncondensible gas (N$_2+$CO$_2$, $i=\mathrm{n}$).
The density ratio of the condensible and noncondensible gases, $\alpha_{\rm v} \equiv \rho_{\rm v}/\rho_{\rm n}$, is related to the temperature as,
\begin{eqnarray}
 &&\frac{{\rm d}\ln \alpha_{\rm v}}{{\rm d}\ln T} = \nonumber \\
 &&\frac
  {
  (R/m_{\rm n})({\rm d} \ln \rho_{\rm v}/ {\rm d} \ln T)
  - (c_{V{\rm n}}/m_{\rm n})
  - \alpha_{\rm v} ({\rm d}s_{\rm v}/{\rm d}\ln T)
  }{
  \alpha_{\rm v}L + R/m_{\rm n}
  },
\label{eq:alpha-temperature}
\end{eqnarray}
where $R\approx 8.3\, {\rm J}\,{\rm mol}^{-1}\,{\rm K}^{-1}$ is the universal gas constant, $m_{\rm n}$ is the mean molar mass of the noncondensible gases, $c_{V{\rm n}}$ is the specific heat at constant volume of the noncondensible gases, $s_{\rm v}$ is the specific entropy of the condensible gas, and $L$ is the latent heat of the condensible component of the evaporation or sublimation.

To compute the pseudo-moist adiabatic lapse rate, the troposphere is assumed to be saturated with H$_2$O vapor, so that the amount of H$_2$O at each altitude is determined by the saturated vapor pressure corresponding to the local temperature.
However, since this assumption does not reproduce the present Earth’s climate, a common approach is to prescribe a sub-saturated relative-humidity (RH) profile when computing the greenhouse effect and radiative transfer \citep[e.g.,][]{Manabe+Wetherald1967,2016A&A...592A..36G}.
While the assumption of $\mathrm{RH}=1$, i.e., saturation of H$_2$O, may be justified for warmer conditions \citep[e.g.,][]{Kasting+Ackerman1986Sci...234.1383K,1988Icar...74..472K,Ramirez+2014AsBio..14..714R,Seeley+Wordsworth2023PSJ.....4...34S}, all the parameters we calculate do not necessarily apply to cases where this assumption is valid.
\edit1{We nevertheless compute the pseudo-moist adiabat assuming $\mathrm{RH}=1$ throughout, and discuss this limitation in Section~\ref{subsec:discussion:limitations}.}

The adiabatic lapse rate is calculated upward from the planetary surface, and the altitude at which the atmospheric temperature reaches $200$ K is defined as the tropopause.
The cold trap for H$_2$O is set at this altitude, where the atmosphere is connected to an isothermal stratosphere.
The H$_2$O mixing ratio in the stratosphere is assumed to be equal to that at the cold trap.
These assumptions are adopted for simplicity and have little influence on the surface temperatures, which are our primary focus.

\subsection{Numerical framework}

We use the \texttt{CLIMA} module contained in a public code, \texttt{Atmos}\footnote{https://github.com/VirtualPlanetaryLaboratory/atmos} \citep{Arney+2016AsBio..16..873A}.
The \texttt{CLIMA} module was originally developed by \citet{2013ApJ...765..131K}, following the formulation of \citet{1988Icar...74..472K}.
We do not describe the entire \texttt{CLIMA} model here, but we refer to several essential concepts in the following sections.

\subsubsection{Procedure to search for temperature structure}
\label{sec:proc-search-temp}

In the present study, we use a simplified method\footnote{This method is called the ``inverse calculation'' in the \texttt{Atmos} code.} to determine the atmospheric temperature structure without calculating the radiative--convective equilibrium at each layer. 
For each combination of \edit1{the surface partial pressures of} $p_{\mathrm{N_2}}$ and $p_{\mathrm{CO_2}}$, the temperature profile is obtained through the following procedure. 
The steps (1)--(4) are repeated in 1~K increments for surface temperatures ranging from 200~K to 600~K:
(1) Set the planetary surface temperature (and the corresponding saturated vapor pressure of H$_2$O), and construct a pressure grid from the surface to the TOA ($10^{-6}$~bar).
(2) Compute the temperature profile and the H$_2$O vapor pressure along the adiabatic lapse rate from the surface upward.  
(3) When the atmospheric temperature reaches the assumed stratospheric temperature (200~K), define that altitude as the tropopause and the cold trap of H$_2$O.
(4) Perform radiative transfer calculations using the two-stream approximation for both solar and planetary thermal radiation throughout the atmosphere, and evaluate the radiative flux at each altitude (see also Section~\ref{sec:radiative-transfer}).
The surface temperature and associated temperature profile that yields the smallest net radiative flux at the TOA is adopted as the feasible solution.
\edit1{We adopt this grid-based procedure rather than a root-finding method, because the solution is not guaranteed to be unique, and because each inverse calculation is computationally inexpensive.
We note, however, that within the parameter range explored here, the solution was unique in practice.}

\subsubsection{Model parameters}
\label{sec:model-parameters}

To focus on a planet resembling Earth in our solar system, we fix the following parameters based on the properties of the Sun and Earth.
The host star spectrum is assumed to be identical to that of the present Sun.
The zenith angle of the incident stellar radiation is set to $\pi/3$.
The solar constant is taken as $0.8$ times the modern Earth's value to prevent the surface temperature from becoming too high under the assumption of ${\rm RH}=1$.
\edit1{We adopt a surface albedo of 0.32, following \citet{2013ApJ...765..131K} and \citet{2025PSJ.....6..212L}. 
This value is also broadly consistent with the modern-Earth benchmark for $p_{\mathrm{N_2}} = 0.8$ bar and $p_{\mathrm{CO_2}} = 0.003$ bar in the \texttt{CLIMA} framework.}
The planetary radius and surface gravity are identical to those of Earth, and these values are used to define the spatial grid for the radiative transfer calculations.

We then explore how the surface temperature depends on the atmospheric composition by performing a parameter survey with \edit1{the surface partial pressures of N$_2$ and CO$_2$,} $p_{{\rm N}2}$ and $p_{{\rm CO}2}$, as free parameters.
For $p_{{\mathrm N}_2}$, the range $10^{-3}-10^{2}$~bar is divided into $16$ intervals with equal logarithmic spacing.
And for $p_{\mathrm{CO}_2}$, the range $10^{-4}-10^{2}$~bar is divided into $19$ intervals with equal logarithmic spacing.
The temperature structure is investigated for $304$ combinations of $p_{{\mathrm N}_2}$ and $p_{\mathrm{CO}_2}$.

As described in Section \ref{sec:atmosph-struct}, we assume that the troposphere is saturated with H$_2$O vapor (i.e., ${\rm RH}=1$).
We refer to this configuration as the fiducial case.
To isolate the effects of H$_2$O vapor, we also perform calculations with ${\rm RH}=0$, referred to as the dry-atmosphere case.

\subsubsection{Modification of \texttt{CLIMA}}
\label{sec:modest-optim-clima}

We modified \texttt{CLIMA} with several adjustments to ensure numerical stability and physical consistency. 
One major issue arose in the calculation of the H$_2$O mixing ratio: under high H$_2$O conditions, the original code could produce values exceeding unity, causing the simulation to fail. 
We traced this to the formulation of the mixing ratio and revised the algorithm to avoid unphysical values. 
This modification successfully prevented breakdowns across the full parameter range explored.

We also made improvements to the calculation of the adiabatic lapse rate. 
In computing the specific heats of the condensible and noncondensible components, we corrected inconsistencies in the molecular species included. 
Additionally, some coefficients used in the calculation of moist adiabatic lapse rate were corrected to follow the formulation of \citet{1988Icar...74..472K}.

\edit1{We do not employ a variable RH parameterization in the present calculations.
In the original \texttt{CLIMA} setup, arbitrary RH can be prescribed in the radiative-transfer calculation, whereas the lapse rate is computed assuming RH$=1$.
Since a lapse-rate formulation for arbitrary RH is currently lacking, using RH $\neq 1$ would introduce an inconsistency between the lapse-rate and radiative-transfer calculations.}

\subsubsection{Radiative transfer}
\label{sec:radiative-transfer}

We use the radiative transfer module of \texttt{CLIMA} as the basis of our calculations.  
\texttt{CLIMA} accounts for two types of radiation propagating through the atmosphere: solar radiation and planetary thermal radiation.  
Solar radiation is divided into 39 spectral intervals spanning $0.25$--$5.4$~$\mu$m.\footnote{We used a updated radiative-transfer module where the number of visible wavelength bins from 38 to 39 to include CO absorption \citep{2025PSJ.....6..212L}.}  
Planetary thermal radiation is divided into 55 spectral intervals covering $0.7$--$500$~$\mu$m.

The transmissivity of each gas is computed using the correlated-$k$ method \citep{1999JQSRT..62..109K}.  
We adopt the HITEMP 2010 database for H$_2$O absorption and the HITRAN 2008 database for CO$_2$ absorption.  
The absorption coefficients depend on both temperature and pressure, and the pressure dependence represents the effect of pressure broadening.
\edit1{
Water vapor continuum, including both the H$_2$O--H$_2$O and H$_2$O--dry component contributions, is included using the BPS water continuum formulation \citep{2011JGRD..11620302P,2013ApJ...765..131K}.
}
A caveat associated with this treatment is discussed in Section~\ref{subsec:discussion:limitations}.

This radiative-transfer model assumes that absorption by different gaseous species is independent.
Under this assumption, the total atmospheric transmissivity is obtained by multiplying the transmissivities of individual species.
Absorption by the infrared-inactive gas N$_2$ is neglected, whereas its contributions to Rayleigh scattering and pressure broadening are fully accounted for.
Rayleigh scattering by all species is also included in the radiative transfer calculations.

All radiative transfer calculations assume a plane-parallel atmosphere and are performed in a steady-state, one-dimensional vertical column.  
Radiative fluxes are computed using the two-stream approximation developed by \citet{1989JGR....9416287T}.

\section{Results} 
\label{sec:results}

We present the results of our parameter survey, which examines how the climate responds to increasing $p_{\rm N_2}$ for each prescribed value of $p_{\rm CO_2}$.
We first show the response of the surface temperature with increasing $p_\mathrm{N_2}$ and classify the resulting climate response as warming, cooling, or \edit1{neutral} within the explored $p_{\rm N_2}$, $p_{\rm CO_2}$ parameter space in Section~\ref{subsec:results:surface}.
We then propose a classification of the vertical temperature profile into two regimes by using the empirical criterion in Section~\ref{sec:class-temp-prof}.
This framework allows us to identify the role of H$_2$O in shaping the climate response.
Finally, we elucidate the physical mechanisms behind the warming response in Section~\ref{sec:origin-warming-trend} and the cooling response in Section~\ref{sec:origin-cooling-trend} by comparing detailed vertical profiles of several atmospheric quantities between the wet-atmosphere (fiducial) case and the dry-atmosphere case.

\subsection{Response of surface temperature}
\label{subsec:results:surface}

\begin{figure*}[tb!] \centering
 \includegraphics[width=\textwidth]{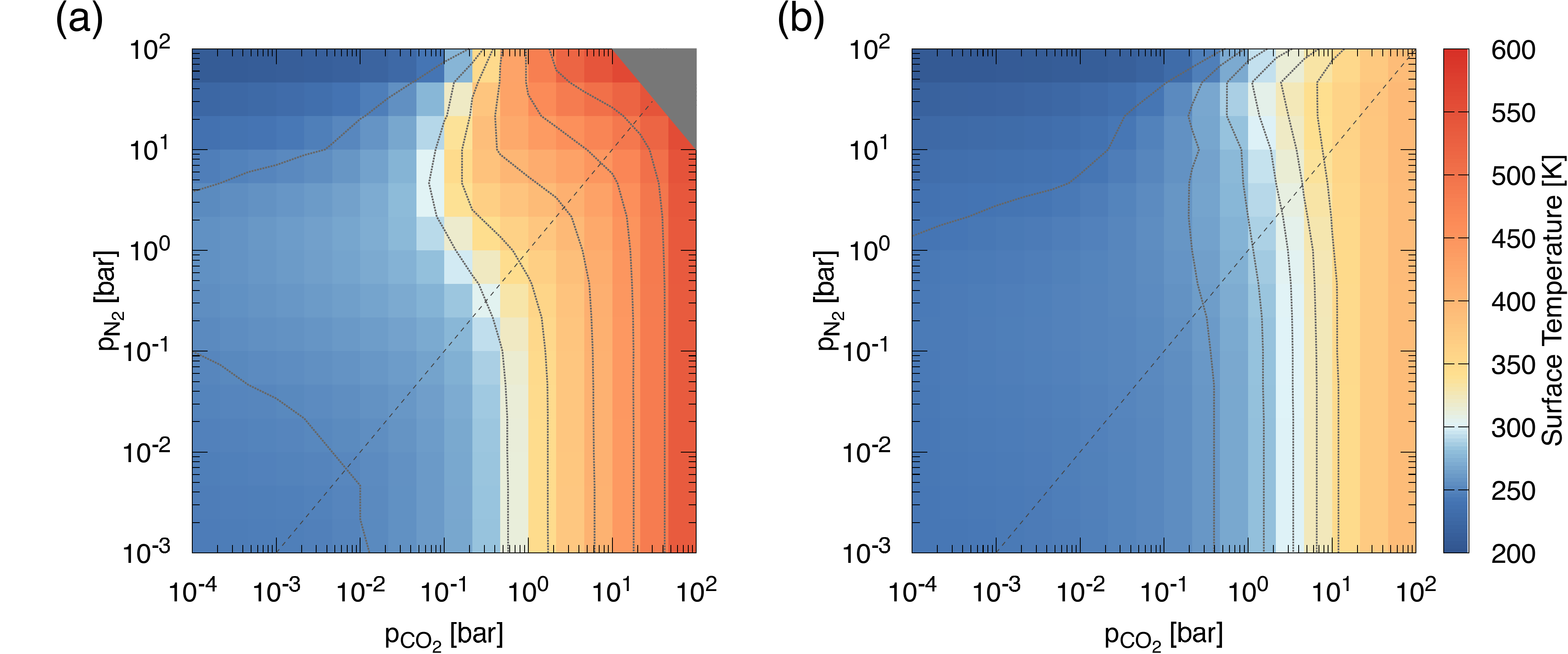}
 \caption{
 Color contour of the surface temperature as a function of $p_{{\rm CO}_2}$ and $p_{{\rm N}_2}$ for (a) the fiducial case and (b) the dry-atmosphere case.
 Dotted lines indicate isotherms, with $T = 250, 300, 350, 400, 450$, and $500$~K in panel (a) and $T = 240, 260, 280, 300, 320$, and $340$~K in panel (b).
 The diagonal dashed line indicates $p_{{\rm CO}_2}=p_{{\rm N}_2}$.
 The gray-shaded region in the upper right of panel (a) denotes the part of parameter space where no consistent atmospheric structure was obtained within the surface temperature range of $200$--$600$~K.
 }
 \label{Fig:SrfTempMap}
\end{figure*}

We found that the climate response to increasing $p_{\rm N_2}$ falls into three categories: \edit1{warming, cooling, or neutral response.}
Figure~\ref{Fig:SrfTempMap} shows color contours of the surface temperature as a function of $p_{\rm CO_2}$ and $p_{\rm N_2}$ for the fiducial case (panel (a)) and the dry-atmosphere case (panel (b)). 
A negative (positive) slope of the contour lines in Figure~\ref{Fig:SrfTempMap} indicates warming (cooling) in response to an increase in $p_{\rm N_2}$. 
Nearly vertical contour lines indicate \edit1{a neutral surface-temperature response to increasing $p_{\rm N_2}$.}

In the fiducial case (panel (a) of Figure~\ref{Fig:SrfTempMap}),
\edit1{the surface-temperature response to increasing $p_{\rm N_2}$ is initially neutral at low $p_{\rm N_2}$, then becomes warming, and eventually cooling as $p_{\rm N_2}$ becomes large.}
In regimes where $p_{\rm CO_2} > p_{\rm N_2}$, the surface temperature is almost insensitive to variations in $p_{\rm N_2}$.
When $p_{\rm N_2}$ approaches $p_{\rm CO_2}$, a warming effect attributable to N$_2$ appears.
At higher values of $p_{\rm N_2}$, the surface temperature begins to decrease.
The value of $p_{\rm N_2}$ at which the surface temperature switches from warming to cooling depends on $p_{\rm CO_2}$.
We revisit the origins of these warming and cooling trends in Sections~\ref{sec:origin-warming-trend} and \ref{sec:origin-cooling-trend}.

To demonstrate the importance of H$_2$O, we also performed a control experiment for the dry-atmosphere case.
The calculations for the dry-atmosphere case adopt the same settings as the fiducial case, except for the absence of H$_2$O.

In the dry-atmosphere case (panel (b) of Figure~\ref{Fig:SrfTempMap}), the warming trend seen in the fiducial case is largely suppressed.
Instead, the surface temperature \edit1{is neutral} over the parameter range where warming appears in the fiducial case (panel (a) of Figure~\ref{Fig:SrfTempMap}).
These results challenge the conventional explanation based on the dominance of pressure broadening \citep{2009NatGe...2..891G,2013ApJ...778..154W}, because that explanation would predict a warming trend even in the absence of H$_2$O.

\subsection{Classification of temperature profiles}
\label{sec:class-temp-prof}

To understand the surface temperature response to increasing $p_{\rm N_2}$, we now take a closer look at the vertical temperature structure.
For this purpose, we classify the vertical temperature structure into two regimes by using an empirical criterion that represents the importance of H$_2$O in controlling the adiabatic lapse rate.
One is the latent–heat–dominated regime, in which the condensation of H$_2$O plays a role in determining the adiabatic lapse rate.
The other is the specific–heat–dominated regime, in which H$_2$O is less important and the lapse rate is nearly identical to the dry adiabatic lapse rate.

\begin{figure*}[tb!] \centering
\includegraphics[width=\textwidth,height=0.9\textheight,keepaspectratio]{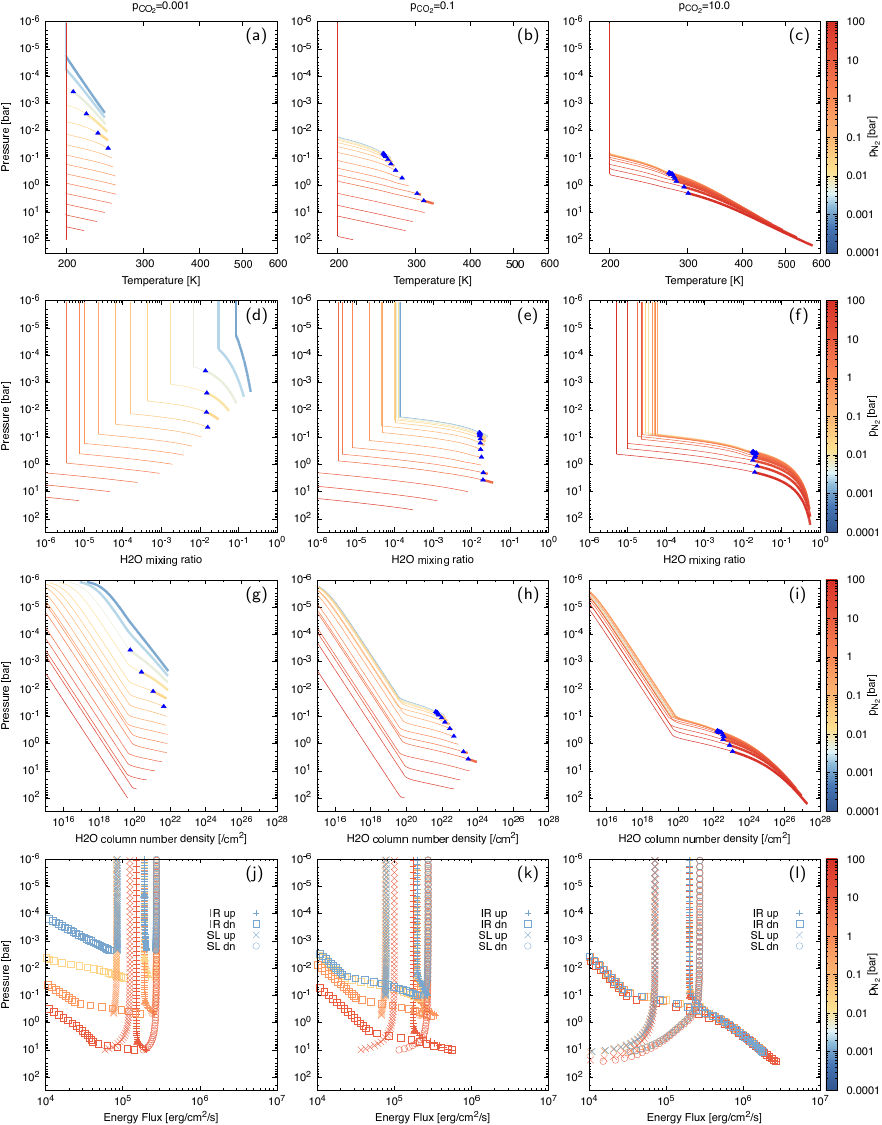}
 \caption{
 Vertical profiles of various atmospheric quantities in the fiducial case.
 Line colors represent $p_{{\rm N}_2}$.
 From left to right, the panels correspond to $p_{{\rm CO}_2} = 10^{-3},\ 10^{-1}$, and $10$ bar, respectively.
 The blue triangular symbol in panels (a)--(i) denotes the boundary between regions governed by the latent-heat-dominated and the specific-heat-dominated lapse rate ($\mathcal{X}=\mathcal{X}_{\rm ref}$).
 (a)--(c): Temperature profiles.
 Thick lines indicate regions following the latent-heat-dominated adiabatic lapse rate, while thin lines correspond to the specific-heat-dominated lapse rate.
 (d)--(f): H$_2$O mixing ratios.
 (g)--(i): Column number densities of H$_2$O.
 (j)--(l): Absolute values of the energy flux.
 The total flux consists of four components: infrared (IR) and solar (SL) radiation, each in upward and downward directions.
 Refer to the figure legend for the correspondence between the markers and the components.
 The triangular symbol denotes the location of the tropopause.
 For clarity, the number of lines shown in panels (j)--(l) is reduced from the actual number of model cases.
}
 \label{Fig:VerticalProfileFiducial}
\end{figure*}

Panels (a)--(c) of Figure~\ref{Fig:VerticalProfileFiducial} show how the temperature gradient varies with $p_{\rm N_2}$ and $p_{\rm CO_2}$.
We here consider three representative values of $p_{\rm CO_2}$: $10^{-3}$, $10^{-1}$, and $10$~bar. 
These cases correspond to low, intermediate, and high CO$_2$ levels.
The temperature in the troposphere is governed by $\Gamma^{-1}_{\rm moist}$ (Equation~(\ref{eq:adiabatic-lapse-rate})).
The upper troposphere is connected to the stratosphere whose temperature is fixed at $200$~K.

\begin{figure*}[tb!]\centering
\includegraphics[width=\textwidth,height=0.9\textheight,keepaspectratio]{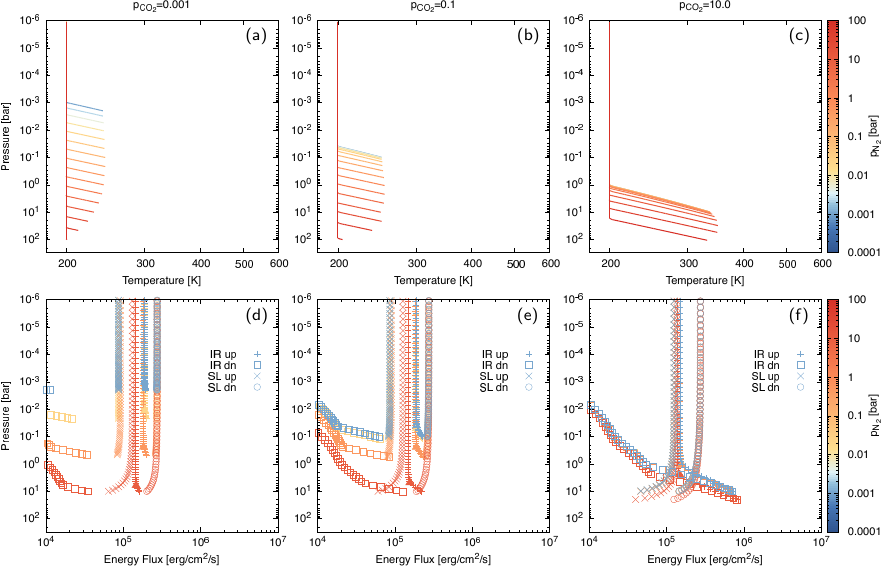}
 \caption{
 A similar set of plots to those in Figure~\ref{Fig:VerticalProfileFiducial}, but for the dry-atmosphere case.
 (a)--(c): Temparature profiles.
 (d)--(f): Absolute values of the energy flux.
 }
 \label{Fig:VerticalProfileDry}
\end{figure*}

The variation in $\Gamma^{-1}_{\rm moist}$ in Figure~\ref{Fig:VerticalProfileFiducial} can be understood as a result of the combination of two limiting cases of the adiabatic lapse rate: one corresponding to the inverse of the dry adiabatic lapse rate, and the other governed by the saturation vapor pressure curve of an ideal gas described by the Clausius–Clapeyron (CC) relation (see Appendix~\ref{sec:effect-trans-adiab} for more detail).
These limiting cases are expressed as
\begin{eqnarray}
   \Gamma^{-1}_{\rm dry} \equiv \left(\frac{{\rm d}\ln p}{{\rm d}\ln T}\right)_{\rm dry} = \frac{c_{p{\rm n}}}{R},
\label{eq:gamma_dry}
\end{eqnarray}
and
\begin{eqnarray}
   \Gamma^{-1}_{\rm CC} \equiv \left(\frac{{\rm d}\ln p_{\rm v}}{{\rm d}\ln T}\right)_{\rm CC} = \frac{m_{\rm v}L}{RT},
\label{eq:gamma_CC}
\end{eqnarray}
where $c_{p{\rm n}}$ is the specific heat at constant pressure of the noncondensible gases, and $m_{\rm v} = 18.0$ is the molar mass of the condensible vapor (H$_2$O).

Which of these two limiting lapse rates is reflected in the atmosphere is primarily determined by the H$_2$O mixing ratio at each latitude.
In H$_2$O-poor conditions, where latent heat release is negligible, $\Gamma^{-1}_{\rm moist}$ approaches $\Gamma^{-1}_{\rm dry}$ (a shallower slope), which is determined by the specific heat of the noncondensible component.
In contrast, in H$_2$O-rich conditions, latent heating becomes significant, and $\Gamma^{-1}_{\rm moist}$ increases toward $\Gamma^{-1}_{\rm CC}$ (a steeper slope).
This behavior of $\Gamma^{-1}_{\rm dry} < \Gamma^{-1}_{\rm CC}$ is due to the energy required for the phase change of the condensible vapor (the latent heat).
To capture the underlying physical distinctions, we refer to these regimes as the specific-heat-dominated regime and the latent-heat-dominated regime, respectively.

To quantify the prevailing regime in an approximate manner, we define the dimensionless number $\mathcal{X}$ as
\begin{eqnarray}
 \mathcal{X} = \left(\frac{p_{\rm v}}{p_{\rm n}} \right)
  \left(\frac{m_{\rm v}L}{RT}\right).
 \label{eq:transition}
\end{eqnarray}
Using $L \approx 2200\,{\rm J}\,{\rm K}^{-1}$ as a representative value, Equation~(\ref{eq:transition}) becomes
\begin{eqnarray}
 \mathcal{X} \approx 1.6 \times 10\left(\frac{p_{\rm v}}{p_{\rm n}} \right)\left(\frac{T}{300 \, {\rm K}}\right)^{-1}.
 \label{eq:transition2}
\end{eqnarray}

By comparing $\mathcal{X}$ with the empirical threshold $\mathcal{X}_{\rm ref}=0.3$, we infer which regime is more likely to apply (see Appendix~\ref{sec:effect-trans-adiab}).
In Figure~\ref{Fig:VerticalProfileFiducial}, the thick (thin) lines in panels (a)--(c) indicate regions with $\mathcal{X} > \mathcal{X}_{\rm ref}$ ($\mathcal{X} < \mathcal{X}_{\rm ref}$), corresponding to the latent–heat–dominated (specific–heat–dominated) regime.

One can confirm that the empirical threshold indeed corresponds to the transition in the slope of $\Gamma^{-1}_{\rm moist}$ in Figure~\ref{Fig:VerticalProfileFiducial}.
We argue that this transition between adiabatic lapse rate regimes produces the warming trend presented in Section~\ref{sec:origin-warming-trend}.

\subsection{The origin of the warming trend}
\label{sec:origin-warming-trend}

To understand the origin of the climate responses observed in Figure~\ref{Fig:VerticalProfileFiducial}, we here assign names to two warming effects associated with H$_2$O: H$_2$O-dilute warming and H$_2$O-load warming.
Each mechanism corresponds to climate changes arising from variations in the H$_2$O mixing ratio, $f_{\rm H_2O}$, and the H$_2$O column number density, $\sigma_{\rm H_2O}$, respectively.
The distinction between H$_2$O-dilute warming and H$_2$O-load warming can also be interpreted in terms of the underlying physical processes, namely the thermodynamic and radiative effects.

H$_2$O-dilute warming refers to the warming trend that arises when an increase in background N$_2$ dilutes the fractional abundance of H$_2$O, thereby modifying the adiabatic lapse rate.
As the dilution shifts the atmosphere from an H$_2$O-rich to an H$_2$O-poor condition, $\Gamma^{-1}_{\rm moist}$ changes from steeper to shallower, as described in Section~\ref{sec:class-temp-prof}.
This change leads to an increase in the surface temperature.
This effect is important when the background gas is initially comparable to or smaller than the H$_2$O mixing ratio. 
Given that $p_{\rm H_2O}\sim 10^{-4}$-$10^{-2}$~bar for the typical temperature range of habitable condition, this warming mechanism is effective when $p_{\rm CO_2} + p_{\rm N_2} \lesssim 10^{-4}$-$10^{-2}$~bar.

H$_2$O-load warming corresponds to the effect traditionally known as the water vapor feedback, in which a slight increase in temperature, initially triggered by mechanisms such as pressure broadening, raises the saturation vapor pressure. 
This leads to an increase in atmospheric H$_2$O, as the enhanced surface temperature promotes stronger H$_2$O loading from the ocean. 
The resulting increase in H$_2$O amplifies the greenhouse effect and causes further warming.

For $p_{\rm CO_2} = 10^{-3}$~bar, the warming is attributed to H$_2$O-dilute warming. 
As $p_{\rm N_2}$ increases, $f_{\rm H_2O}$ at the surface decreases significantly, while $\sigma_{\rm H_2O}$ remains nearly constant (panels (d) and (g) of Figure~\ref{Fig:VerticalProfileFiducial}). 
This is accompanied by a reduction in the extent of the latent-heat-dominated region and an decrease in $\Gamma^{-1}_{\rm moist}$, which lead to surface warming (panel (a) of Figure~\ref{Fig:VerticalProfileFiducial}).
These features are consistent with a mechanism where the thermodynamic structure of the atmosphere, rather than radiative opacity, plays the important role.

For $p_{\rm CO_2} = 10$~bar, the warming is explained by H$_2$O-load warming. 
In this case, the surface $f_{\rm H_2 O}$ remains nearly unchanged, but $\sigma_{\rm H_2O}$ increases significantly with $p_{\rm N_2}$ (panels (f) and (i) of Figure~\ref{Fig:VerticalProfileFiducial}). 
The lapse rate in the lower atmosphere stays almost constant, indicating that the latent-heat-dominated region remains unchanged while the surface temperature continues to rise (panel (c) of Figure~\ref{Fig:VerticalProfileFiducial}). 
This pattern suggests that radiative effects associated with increased $\sigma_{\rm H_2O}$ drives the warming process.

The $p_{\rm CO_2} = 10^{-1}$~bar case exhibits an intermediate behavior, showing contributions from both mechanisms. 
As $p_{\rm N_2}$ increases, $f_{\rm H_2O}$ initially decreases but later stabilizes, while $\sigma_{\rm H_2O}$ shows a gradual increase in the warming regime (panels (e) and (h) of Figure~\ref{Fig:VerticalProfileFiducial}). 
The temperature structure reflects a combination of evolving lapse rate and growing radiative absorption (panel (b) of Figure~\ref{Fig:VerticalProfileFiducial}).
These trends indicate a transitional regime where both thermodynamic and radiative processes play roles in shaping the warming.

Although our analysis does not quantitatively constrain the actual contribution of pressure broadening of H$_2$O lines 
to the warming trend, we consider H$_2$O-dilute and H$_2$O-load warming to be plausible major contributors in the fiducial case.
In panels (a)--(c) of Figure~\ref{Fig:VerticalProfileDry}, no significant warming attributable to pressure broadening is observed up to $p_{\rm N_2} \approx 1$~bar, regardless of $p_{\rm CO_2}$ in the dry-atmosphere case.
This suggests that H$_2$O behaves in a similar manner.

\subsection{The origin of the cooling trend}
\label{sec:origin-cooling-trend}

The cooling trend at high $p_{\rm N_2}$ is driven by Rayleigh scattering, as seen in the vertical profiles of the energy flux.

Panels (j)--(l) of Figure~\ref{Fig:VerticalProfileFiducial} and panels (d)--(f) of Figure~\ref{Fig:VerticalProfileDry} show the absolute value of the vertical energy flux in the fiducial and dry-atmosphere cases.
In cases exhibiting cooling, the upward solar flux (labeled ``SL up'' in the figures) at the TOA has a large absolute value, corresponding to a smaller net stellar energy input to the atmosphere.

These panels show that the outward solar flux responsible for the cooling is generated within the atmosphere.
The upward solar flux decreases toward the surface, indicating that it is not set by surface reflection but is produced within the atmospheric column, as expected for Rayleigh scattering.

The cooling might be enhanced by the H$_2$O vapor feedback, in which a decrease in surface temperature reduces the atmospheric H$_2$O loading and weakens the H$_2$O greenhouse effect.
To assess this effect, we compare panels (a)--(b) of the fiducial case (Figure~\ref{Fig:VerticalProfileFiducial}) with those of the dry-atmosphere case (Figure~\ref{Fig:VerticalProfileDry}).
The value of $p_{\rm N_2}$ at which the cooling trend appears is nearly the same in both cases.
However, the magnitude of the temperature decrease is larger in the fiducial case than in the dry-atmosphere case.
This difference indicates that the H$_2$O vapor feedback amplifies the cooling in the fiducial atmosphere.

Another important controlling factor in the formation of the cooling trend is the optical depth of the atmosphere.
A large amount of atmospheric H$_2$O increases the surface temperature and also makes $\Gamma^{-1}_{\rm moist}$ steeper.
This effect elevates the tropopause, which results in a greater amount of H$_2$O at higher altitudes.
Eventually, the downward solar flux is absorbed before reaching the pressure level at which Rayleigh scattering becomes effective, so that Rayleigh scattering cannot contribute to the TOA energy balance.
Because of this mechanism, the cooling trend does not appear in the fiducial case with $p_{\rm CO_2}=10$~bar (panel (c) of Figure~\ref{Fig:VerticalProfileFiducial}).
In this case, Rayleigh scattering modifies the vertical profile of the upward solar flux within the atmosphere, but the upward solar flux at the TOA remains nearly unchanged with increasing $p_{\rm N_2}$ (panel (l) of Figure~\ref{Fig:VerticalProfileFiducial}).
\section{Discussion} \label{sec:discussions}

\subsection{Evaluation of model consistency}
\label{sec:validations}

Our methodological choices, namely, imposing an isothermal stratosphere and selecting solutions by the TOA energy-balance residual, do not have practical consequences for the inferred relation between surface temperature and atmospheric composition.
This conclusion is supported by the vertical energy flux profiles and by the small residual of the TOA energy balance in the adopted solutions.

The isothermal-stratosphere assumption would affect the results if a substantial fraction of the upward planetary infrared flux (IR up) were absorbed above the tropopause, because the stratospheric temperature structure would then influence the infrared energy transport.
In panels (j)--(l) of Figure~\ref{Fig:VerticalProfileFiducial} and panels (d)--(f) of Figure~\ref{Fig:VerticalProfileDry}, the tropopause (triangular symbol) lies above the region where IR up is mainly modified, indicating that the dominant infrared heating and cooling occur at pressures higher than the tropopause pressure.
Thus, the imposed isothermal stratosphere has only a minor impact on our parameter survey.

In the adopted solutions, the TOA energy imbalance is at most $1$\% and is below $0.1$\% in most cases (Appendix~\ref{sec:energy-balance-at}).
No well-balanced solution is found in a limited part of the parameter space, which is shaded in gray in panel (a) of Figure~\ref{Fig:SrfTempMap}.

\subsection{Comparison with previous studies}
\label{subsec:discussion:comparison}

Based on our extensive parameter survey and analysis (Section~\ref{sec:results}), we discuss the causes of the different climate effects reported in previous studies \citep{2009NatGe...2..891G,2013ApJ...778..154W}.
the different climate effects reported in previous studies \citep{2009NatGe...2..891G,2013ApJ...778..154W}.
\citet{2009NatGe...2..891G} showed warming with increasing $p_{\mathrm{N}_2}$ by comparing $p_{\mathrm{N}_2}=0.4$, $0.8$, $1.6$, and $2.4$~bar cases for $p_{\mathrm{CO}_2}=10^{-4}$--$10^{-2}$~bar (their Figure~1a).
\citet{2013ApJ...778..154W} showed both warming and cooling with increasing $p_{\mathrm{N}_2}$ by comparing $p_{\mathrm{N}_2}=0.8$ and $4.0$~bar cases for $p_{\mathrm{CO}_2}=10^{-4}$--$10^{1}$~bar (their Figure~14a).
Here we focus on their $g=9.81\,\mathrm{m\,s^{-2}}$ case as an Earth analog.

\begin{figure}[tbp] \centering
 \includegraphics[width=0.99\columnwidth]{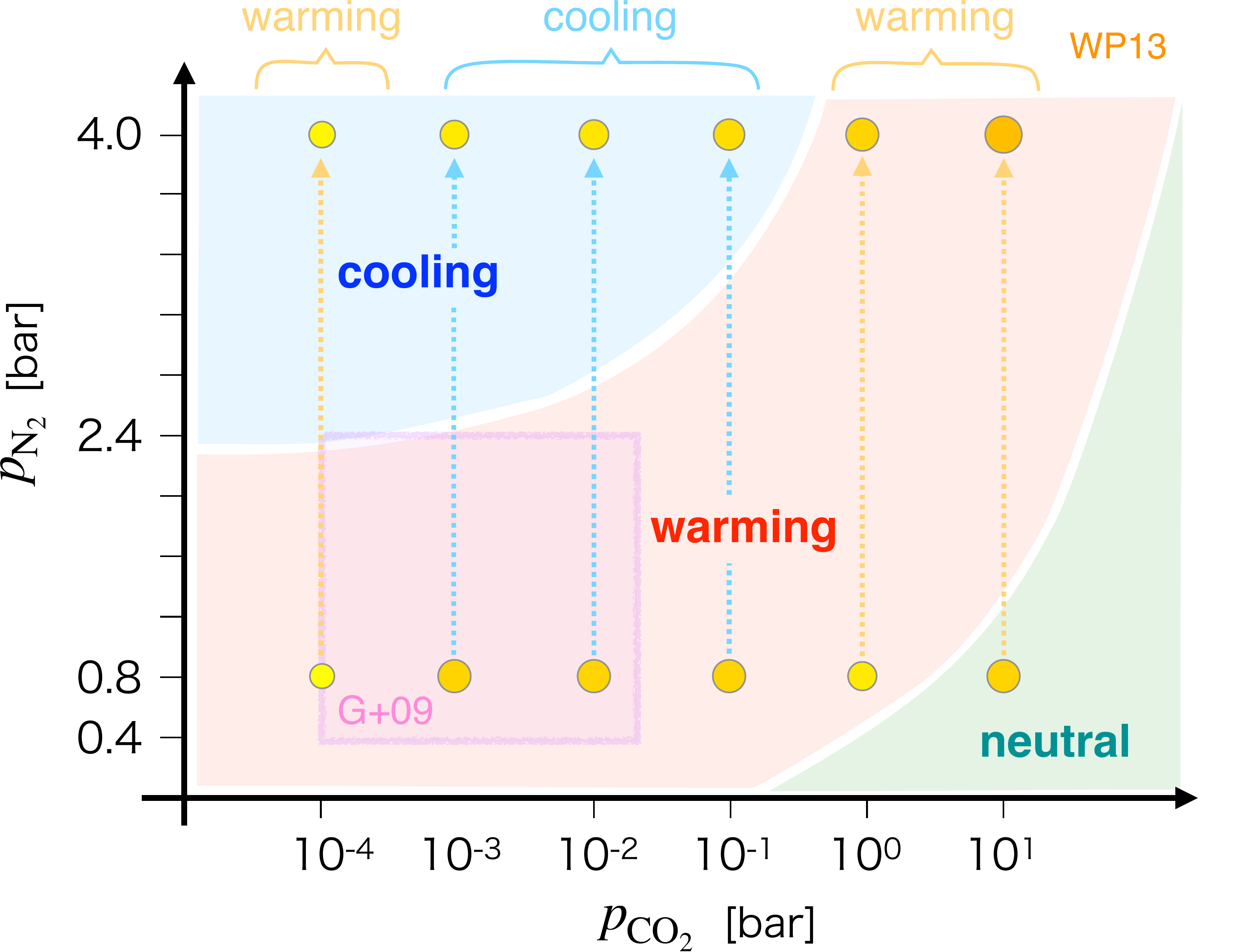}
 \caption{
Schematic summary of the surface temperature response to increasing $p_{\rm N_2}$ on the $p_{\rm CO_2}$--$p_{\rm N_2}$ plane, comparing \citet[][G+09]{2009NatGe...2..891G}, \citet[][WP13]{2013ApJ...778..154W}, and this work.
For \citet{2009NatGe...2..891G}, the pink shaded region denotes the parameter range they explored, within which all cases show warming as $p_{\rm N_2}$ increases.
For \citet{2013ApJ...778..154W}, circles mark the parameter sets shown in their Figure~14a, and the annotation at the top summarizes their reported response.
Circle size and color qualitatively indicate the surface temperature, with larger circles and warmer colors corresponding to higher temperatures.
In \citet{2013ApJ...778..154W}, the sign of the response is inferred from the two data points connected by arrows, which indicate the direction of increasing $p_{\rm N_2}$.
Background shading shows the approximate regimes obtained in this work, namely cooling, warming, and \edit1{neutral surface-temperature response to increasing $p_{\rm N_2}$.}
}
 \label{Fig:schematic}
\end{figure}

Figure~\ref{Fig:schematic} summarizes the surface temperature response to increasing $p_{\mathrm{N}_2}$ in the three studies.
The parameter range explored by \citet{2009NatGe...2..891G} largely falls within the warming regime in our study.
For $p_{\rm CO_2}\geq 1$~bar, the warming trend reported by \citet{2013ApJ...778..154W} is consistent with our results.
For $p_{\rm CO_2}\leq 10^{-1}$~bar, the comparison is less straightforward because their two data points bracket the warming-to-cooling transition in our calculations.
In this case, the apparent trend between the two points can be warming or cooling, depending on how close each point is to the transition and on the local slopes on either side.
This also suggests that resolving climatic impact of increasing $p_{\rm N_2}$ requires sampling $p_{\mathrm{N}_2}$ with finer spacing.

The $p_{\rm N_2}$--$p_{\rm CO_2}$ boundary for the warming-to-cooling transition likely depends on the adopted relative-humidity model, because this transition is sensitive to the strength of the H$_2$O vapor feedback (Sections~\ref{sec:origin-warming-trend} and \ref{sec:origin-cooling-trend}).
\citet{2009NatGe...2..891G} assumed a modern-Earth relative humidity profile \citep{Manabe+Wetherald1967} under a fainter Sun, whereas \citet{2013ApJ...778..154W} assumed RH$=1$ with current-Earth-level insolation.
Accordingly, the reported transition should be compared across studies with these differing humidity assumptions in mind.

We update the interpretation of the surface temperature response to increasing N$_2$, focusing on the mechanism of N$_2$-induced warming.
\citet{2009NatGe...2..891G} attributed the warming mainly to pressure broadening while the pressure effect on the lapse rate and H$_2$O vapor feedback were also noted.
\citet{2013ApJ...778..154W} discussed the balance between pressure broadening and Rayleigh scattering.
Our dry-atmosphere experiment shows that the warming region is much smaller than in the wet (fiducial) case, which implies that the main driver of N$_2$-induced warming is the increase in the H$_2$O vapor column density.

\edit1{
\citet{Keles+2018AsBio..18..116K} reported a qualitatively similar warming-to-cooling transition in a modern-Earth-like atmosphere.
Using a 1D cloud-free coupled climate--photochemistry model for an N$_2$--O$_2$ atmosphere at 1 au, they found that the surface temperature increases with pressure up to $\sim 4$~bar, but decreases at higher pressure as Rayleigh scattering becomes dominant.
}

\edit1{
A direct comparison with our results is nevertheless limited by the structure of their setup.
In their model, N$_2$, O$_2$, and CO$_2$ are fixed as isoprofiles while the total surface pressure is varied.
Increasing pressure therefore changes both the atmospheric mass and the column amount of radiatively active species such as CO$_2$.
This makes it difficult to separate the contributions of pressure broadening, Rayleigh scattering, and H$_2$O feedback.
By contrast, our calculations provide a more idealized framework for isolating these effects and suggest that the behavior identified by \citet{Keles+2018AsBio..18..116K} is not restricted to a modern-Earth setting, but may arise more generally across a wider range of terrestrial surface environments.
}

\edit1{
\citet{2019ApJ...881..120K} provides another useful point of comparison because it also emphasizes the climatic importance of H$_2$O-mediated effects.
They argued that background gases affect climate not only through direct radiative effects, but also through their influence on atmospheric H$_2$O and thermal structure.
In this respect, their interpretation is broadly consistent with ours.
}

\edit1{
The key difference is that \citet{2019ApJ...881..120K} considered H$_2$ as a background gas, which substantially changes the mean molecular weight of the atmosphere and, in turn, modifies how the H$_2$O feedback operates.
As a result, N$_2$-rich and H$_2$-rich atmospheres approach the runaway greenhouse in qualitatively different ways.
Our study instead focuses on atmospheres in which N$_2$ remains the background gas and examines temperate to moderately warm Earth-like climates, showing that increasing $p_{\rm N_2}$ can produce either warming or cooling depending on $p_{\rm CO_2}$.
}

\edit1{
\citet{2022A&A...658A..40C} is also relevant because it shows that background gases can influence climate not only through direct radiative effects associated with H$_2$O line broadening, but also through changes in the adiabatic lapse rate, and thus in the atmospheric temperature structure.
Using a suite of 1D radiative--convective models, they examined the onset of the runaway greenhouse in H$_2$O + N$_2$ atmospheres and focused on the behavior of the OLR near the Simpson--Nakajima limit.
}

\edit1{
A direct comparison between our study and that of \citet{2022A&A...658A..40C} is not straightforward because the two studies were designed to address different questions.
Their study was designed to identify the physical origin of the OLR overshoot in hot, H$_2$O-rich atmospheres.
In contrast, our study examines the surface-temperature response to increasing $p_{\rm N_2}$ across a broader $p_{\rm CO_2}$--$p_{\rm N_2}$ parameter space, spanning temperate to moderately warm climates.
}

\edit1{
Nevertheless, the two studies share an important physical perspective.
In \citet{2022A&A...658A..40C}, the role of background gases appears through a transition from foreign- to self-broadening of H$_2$O absorption lines and through a transition from dry- to moist-adiabatic thermal structure as the atmosphere evolves from an N$_2$-dominated to an H$_2$O-dominated regime.
In our study, related H$_2$O-mediated and lapse-rate-related effects also play a central role in organizing the warming and cooling trends with increasing $p_{\rm N_2}$.
Thus, the two studies emphasize similar physical ingredients, even though those ingredients appear differently because the climatic regimes and diagnostics are different.
}

\subsection{Implications for different terrestrial planets}
\label{subsec:discussion:implications}

Here we discuss implications for different terrestrial planets based on our revised understanding of the greenhouse effects of non-greenhouse gases.
We note that the quantitative effects of non-greenhouse gases are likely to depend on several factors that are fixed in this study (Section \ref{subsec:discussion:limitations}), and thus future studies dedicated to individual planets are necessary to draw quantitative conclusions.

The climate effect of potentially-high $p_{{\mathrm N}_2}$ on early Earth is dependent on its absolute value and Earth's climate state.
\cite{2009NatGe...2..891G} showed that increasing $p_{{\mathrm N}_2}$ up to $1.6$ bar leads to warming. 
However, our fiducial model suggests that further increase in $p_{{\mathrm N}_2}$ causes cooling (panel (a) of Figure \ref{Fig:SrfTempMap}), provided that our results are still applicable for the Archean Earth condition where the sun is fainter and additional CH$_4$ is present. 
Such cooling may be relevant to Hadean Earth and its climate evolution, which potentially started from a dense N$_2$-rich atmosphere \citep{Kurokawa+2022GGG....2310295K}. 
Furthermore, our results for the dry-atmosphere case (panel (b) of Figure \ref{Fig:SrfTempMap}) suggests limited N$_2$ warming on early Earth during snowball periods, where lower surface temperature leads to lower atmospheric water vapor content.

Similarly, whether $p_{{\mathrm N}_2}$ and other non-greenhouse gases led to warming or cooling on early Mars and Venus is also be dependent on whether their atmospheres are wet or dry. 
Nitrogen isotopic constraints suggest higher $p_{{\mathrm N}_2}$ on early Mars \citep{Jakosky+1994Icar..111..271J,Kurokawa+2018Icar..299..443K,Hu+Thomas2022NatGe..15..106H,Thomas+2023PSJ.....4...41T}. 
Record of fluvial features and hydrated minerals generally suggests that early Mars sustained liquid water, at least intermittently \citep[e.g.,][]{Ehlmann+2016JGRE..121.1927E,Ramirez+Craddock2018NatGe..11..230R}. 
However, this may not mean that its atmosphere was wet as well; for instance, \cite{Kite+2021PNAS..11801959K} proposed an arid scenario for early Mars' atmosphere, in which a warm climate is enabled by high-altitude water ice clouds. 
Because hydrogen isotopic record suggests that the significant water loss predates the ages of fluvial features \citep{Kurokawa+2014E&PSL.394..179K}, the climate of early Mars and the impact of a dense N$_2$-rich atmosphere on it needs to be considered together with the evolution of surface water reservoirs.

CO is another potentially abundant non-greenhouse gas on early Mars \citep{zahnle2008photochemical,Sholes+2017Icar..290...46S,Ueno+2024NatGe..17..503U}. 
Runaway photolysis of CO$_2$ to form CO \citep{zahnle2008photochemical} leads to reduction of CO$_2$ greenhouse effect and cools Mars' climate. 
Under such cold conditions, CO may cause further cooling due to the dominance of its scattering effect. 

The climate state and thus the climate effects of N$_2$ on early Venus are further uncertain, due to limitation in both record and access to it. 
Whereas many studies argue that Venus never experienced a habitable condition \citep{Gillmann+2009,Gillman+2020,Hamano+2013,Turbet+2021}, some studies proposed a habitable early-Venus scenario \citep{Way+2016GeoRL..43.8376W,Way+DelGenio2020JGRE..12506276W}. 
Under the habitable (wet) state assumed, N$_2$ contributes to warming. 
As Venus turned into a desiccated planet, the transition to N$_2$ warming to cooling might have happened.

The climate effect of non-greenhouse gases is likely shifted to warming on planets orbiting M-type stars, which are primary targets for characterizing rocky exoplanet atmospheres and habitability. 
Radiation from host stars shifts to longer wavelengths, which reduces Rayleigh scattering and cooling \citep{2025PSJ.....6..212L}, while enhances warming by pressure broadening and water feedback (Section \ref{subsec:discussion:limitations}). 
Therefore, CO-O$_2$ runaway on these planets \citep{Tian+2014,harman2015abiotic,Hu+2020} may lead to surface warming, provided that $p_{\mathrm{CO}_2}$ is regulated by another mechanism, presumably by carbonate-silicate cycling \citep{1981JGR....86.9776W,1983AmJS..283..641B}.

\subsection{Caveat} 
\label{subsec:discussion:limitations}

Although our conclusions for the roles of four mechanisms on greenhouse and anti-greenhouse effects of non-greenhouse gases will still hold, our quantitative results (e.g., where in the parameter space warming/cooling dominates) are likely dependent on our model assumptions and settings. 
In this study, we fixed the RH model ($\mathrm{RH} = 1$), neglected \edit1{N$_2$--N$_2$} CIA, and used the inverse method.
We also fixed the host-stellar type and the stellar radiation flux at the TOA to be the sun and the 0.8 times the modern Earth's insolation, respectively.

The RH model assumed in a 1D climate model and its parameter dependence impact the climate effects of non-greenhouse gases. 
While Many 1D climate models assumes the prescription of \cite{Manabe+Wetherald1967}, detailed analysis of global satellite observations over Earth's oceans found spatially-varying RH profiles \citep{Abraham+Goldblatt2022JAtS...79.2243A}. 
Moreover, several previous studies on 1D climate modeling assumed positive feedback between the surface temperature and the RH profile \citep{Kasting+Ackerman1986Sci...234.1383K,Goldblatt+2013NatGe...6..661G,Ramirez+2014AsBio..14..714R}. 
If such RH models are adapted, the response of the RH to the surface temperature will amplify the water vapor feedback, affecting our results quantitatively.

\edit1{
Another source of quantitative uncertainty is the treatment of the water vapor continuum.
The water vapor continuum is expected to be an important opacity source for planetary radiation, especially in warming cases with abundant atmospheric H$_2$O.
Indeed, \citet{2016ApJ...826..222Y} showed that the choice of the water vapor continuum model produces quantitatively significant differences in the outgoing longwave radiation.
In the present study, however, our primary interest is in the qualitative mechanism of the climate response, and we therefore did not investigate the dependence on the continuum model choice.
Accordingly, caution is required in interpreting the quantitative aspects of our results.
}

N$_2$\edit1{--N$_2$ CIA} is not included in our model. 
\edit1{
\citet{2022A&A...658A..40C} showed, for a setup similar to our fiducial case, that including N$_2$--N$_2$ CIA changes the planetary radiation by less than $1\%$ even at $p_{\mathrm{N}_2} = 10$ bar.
Therefore, for our fiducial calculations up to $p_{\mathrm{N}_2} \sim 10$ bar, the effect of N$_2$--N$_2$ CIA is expected to be negligible.
}

\edit1{
N$_2$--N$_2$ CIA is expected to become more important at higher N$_2$ pressures.
Its radiative effect should be particularly important when H$_2$O absorption is weak, for example in drier atmospheres \citep[e.g.,][]{2024JQSRT.32809172S}.
At $p_{\mathrm{N}_2} > 10$ bar and $p_{\mathrm{CO}_2} < 0.1$ bar, where Rayleigh-scattering-induced cooling is strong and atmospheric H$_2$O is scarce, the cooling found in our calculations may be somewhat overestimated.
}

As suggested in Section~\ref{subsec:discussion:implications}, the stellar type can shift the balance between warming and cooling.
Because the Rayleigh scattering cross section is inversely proportional to the fourth power of wavelength \citep{catling2017atmospheric}, its cooling effect becomes less prominent for later-type stars, which emit radiation at longer wavelengths \citep{2025PSJ.....6..212L}.

In addition, variations in the incoming stellar irradiation flux influence the strength of the H$_2$O vapor feedback.
For instance, adopting a lower irradiation level leads to generally lower surface temperatures and reduced atmospheric H$_2$O, thereby weakening the effects of the vapor feedback.

\section{Conclusions}
\label{sec:conclusion}

In this study, we revisited how increases in non-greenhouse gases influence planetary climate, focusing on the surface temperature and the vertical thermal and H$_2$O vapor structures.
We conducted numerical experiments in which the partial pressure of N$_2$ was increased as a proxy for non-greenhouse gases in an N$_2$--CO$_2$--H$_2$O atmosphere.

We showed that the climate response to increasing N$_2$ depends on the CO$_2$ partial pressure, $p_{\rm CO_2}$, even though the overall behavior transitions from \edit1{neutral} to warming and eventually to cooling across a wide range of $p_{\rm CO_2}$.

In the warming regime, under low $p_{\rm CO_2}$ conditions, dilution of atmospheric H$_2$O by additional N$_2$ modifies the adiabatic lapse rate and warms the surface.
Under high $p_{\rm CO_2}$ conditions, enhanced H$_2$O loading driven by water vapor feedback produces additional surface warming.
We confirmed that these two H$_2$O-related warming effects play an important role by conducting numerical experiments, which show that the warming trend induced by increasing N$_2$ almost disappears in the absence of atmospheric H$_2$O (the dry-atmosphere case).

In the cooling regime, Rayleigh scattering plays a key role by reducing the energy supplied to the surface under low to intermediate $p_{\rm CO_2}$ conditions.
The water vapor feedback further strengthens the cooling by decreasing the H$_2$O column density and weakening the greenhouse effect.
Under high-$p_{\rm CO_2}$ conditions, when the atmospheric H$_2$O loading becomes large, the tropopause rises and the optical depth of the atmosphere increases.
In this situation, the cooling trend disappears because the absorption of downward solar flux by H$_2$O, rather than Rayleigh scattering, controls the TOA energy balance.

These findings clarify the multiple physical origins of non-greenhouse gas driven climate responses and provide a predictive framework for estimating both the direction and the magnitude of such responses in the atmospheres of Earth-like planets.

\begin{acknowledgments}
We thank the anonymous referee for careful reading and constructive comments that helped us improve the manuscript.
We sincerely appreciate the fruitful discussions with Dr. Colin Goldblatt, which greatly contributed to the progress of this research. 
This work was supported by JSPS KAKENHI Grant Number 21K13983, 22H05150, 20KK0080, 22H01290, 21H04514, 23K22561, and 25K01062.
Numerical computations were in part carried out on PC cluster at Center for Computational Astrophysics, National Astronomical Observatory of Japan.
\end{acknowledgments}

\vspace{5mm}
% \software{astropy \citep{2013A&A...558A..33A,2018AJ....156..123A},  
%          Cloudy \citep{2013RMxAA..49..137F}, 
%          Source Extractor \citep{1996A&AS..117..393B}
%          }

\appendix
\section{The effective transition of the adiabatic lapse rate}
\label{sec:effect-trans-adiab}

We introduce a criterion for distinguishing between the latent-heat-dominated and specific-heat-dominated regimes of the adiabatic lapse rate.
Assuming an ideal gas, the derivative $\mathrm{d}\ln p_{\rm v}/\mathrm{d}\ln T$ follows the Clausius–Clapeyron relation.
Under this assumption, the inverse of the pseudo-moist adiabatic lapse rate (Equation~(\ref{eq:adiabatic-lapse-rate})) can be written as
\begin{eqnarray}
 \Gamma^{-1}_{\rm moist} =
  \Gamma^{-1}_{\rm CC}
  \left\{1-
   \frac{1-
   (\Gamma^{-1}_{\rm dry}/\Gamma^{-1}_{\rm CC})/[1+(p_{\rm v}/p_{\rm n})]}
   {1+(p_{\rm v}/p_{\rm n}) \Gamma^{-1}_{\rm CC}}
  \right\},
\label{eq:MoistAdiabatAppendix}
\end{eqnarray}
where $\Gamma^{-1}_{\rm CC}$ and $\Gamma^{-1}_{\rm dry}$ are defined in Equations~(\ref{eq:gamma_CC}) and~(\ref{eq:gamma_dry}).
The term $(p_{\rm v}/p_{\rm n}) \Gamma^{-1}_{\rm CC}$ in the denominator corresponds to the dimensionless number $\mathcal{X}$, which we introduced to characterize the transition between different adiabatic lapse rate regimes (Equation~(\ref{eq:transition})).
Equation~(\ref{eq:MoistAdiabatAppendix}) approaches $\Gamma^{-1}_{\rm CC}$ and $\Gamma^{-1}_{\rm dry}$ in the respective limits of $\mathcal{X} \gg 1$ and $\mathcal{X} \ll 1$.
By inserting representative values, we obtain Equation~(\ref{eq:transition2}), which shows that $\mathcal{X}$ is determined solely by the partial pressure ratio of the condensible to noncondensible components and the temperature at each pressure.

We found that a reference value of $\mathcal{X}_{\rm ref}=0.3$ can be used as an empirical threshold.
For $\mathcal{X} < 0.3$, $\Gamma^{-1}_{\rm moist}$ rapidly approaches to $\Gamma^{-1}_{\rm dry}$.
Therefore, we classify the region with $\mathcal{X} < 0.3$ as the specific-heat-dominated regime, and the complementary region with $\mathcal{X} > 0.3$ as the latent-heat-dominated regime.

\section{Energy balance at the top of the atmosphere}
\label{sec:energy-balance-at}

\begin{figure*}[tb!] \centering
 \includegraphics[width=\textwidth]{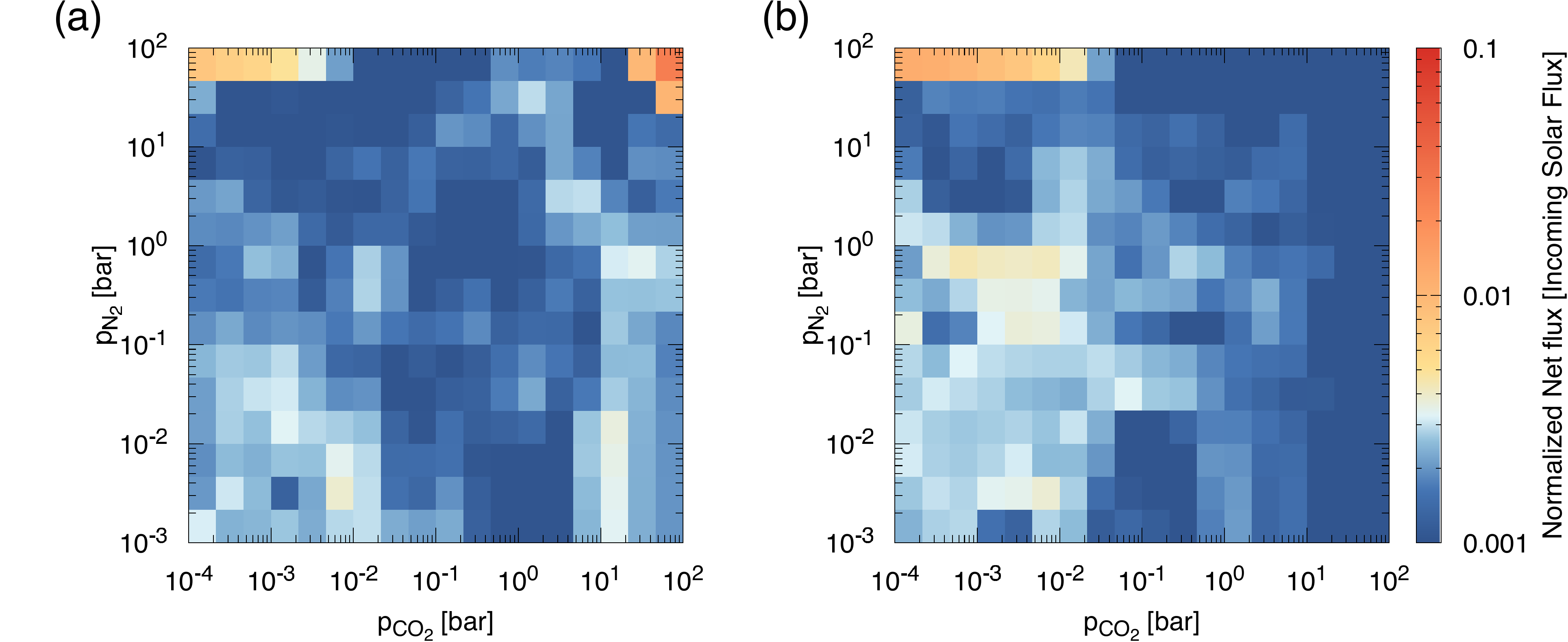}
 \caption{
TOA net energy flux normalized by the incoming solar flux.
(a) fiducial case, (b) dry-atmosphere case.
 }
 \label{Fig:NetFluxMap}
\end{figure*}

Figure~\ref{Fig:NetFluxMap} shows the TOA net energy flux for the fiducial case (panel~(a)) and the dry-atmosphere case (panel~(b)).
As described in Section~\ref{sec:proc-search-temp}, we adopt the equilibrium structure that yields the surface temperature at which the TOA net energy flux is closest to zero.

Since the residual flux is typically $\sim 0.1$\% over a wide range of parameters, and reaches $\sim 1$\% in some regions, the TOA energy balance is well satisfied.
A larger deviation of $\sim 10$\% appears in the upper-right portion of the fiducial case, where no physically plausible solution is obtained.
This region is shaded in gray in panel (a) of Figure~\ref{Fig:SrfTempMap}.

%% ------------------------------ %%
\bibliography{reflist}{}

@ARTICLE{2016ApJ...826..222Y,
       author = {{Yang}, Jun and {Leconte}, J{\'e}r{\'e}my and {Wolf}, Eric T. and {Goldblatt}, Colin and {Feldl}, Nicole and {Merlis}, Timothy and {Wang}, Yuwei and {Koll}, Daniel D.~B. and {Ding}, Feng and {Forget}, Fran{\c{c}}ois and {Abbot}, Dorian S.},
        title = "{Differences in Water Vapor Radiative Transfer among 1D Models Can Significantly Affect the Inner Edge of the Habitable Zone}",
      journal = {\apj},
         year = 2016,
        month = aug,
       volume = {826},
       number = {2},
          eid = {222},
        pages = {222},
          doi = {10.3847/0004-637X/826/2/222},
archivePrefix = {arXiv},
       eprint = {1809.01397},
 primaryClass = {astro-ph.EP},
       adsurl = {https://ui.adsabs.harvard.edu/abs/2016ApJ...826..222Y}
}

@ARTICLE{2024JQSRT.32809172S,
       author = {{Serov}, E.~A. and {Galanina}, T.~A. and {Koroleva}, A.~O. and {Makarov}, D.~S. and {Amerkhanov}, I.~S. and {Koshelev}, M.~A. and {Tretyakov}, M. Yu. and {Chistikov}, D.~N. and {Finenko}, A.~A. and {Vigasin}, A.~A.},
        title = "{Continuum absorption in pure N<mml:math altimg=``si97.svg'' display=``inline'' id=``d1e1392''><mml:msub><mml:mrow></mml:mrow><mml:mrow><mml:mn>2</mml:mn></mml:mrow></mml:msub></mml:math> gas and in its mixture with Ar}",
      journal = {\jqsrt},
         year = 2024,
        month = dec,
       volume = {328},
          eid = {109172},
        pages = {109172},
          doi = {10.1016/j.jqsrt.2024.109172},
       adsurl = {https://ui.adsabs.harvard.edu/abs/2024JQSRT.32809172S}
}

@ARTICLE{2011JGRD..11620302P,
       author = {{Paynter}, D.~J. and {Ramaswamy}, V.},
        title = "{An assessment of recent water vapor continuum measurements upon longwave and shortwave radiative transfer}",
      journal = {Journal of Geophysical Research (Atmospheres)},
         year = 2011,
        month = oct,
       volume = {116},
       number = {D20},
          eid = {D20302},
        pages = {D20302},
          doi = {10.1029/2010JD015505},
       adsurl = {https://ui.adsabs.harvard.edu/abs/2011JGRD..11620302P}
}

@ARTICLE{2020MNRAS.494..259R,
       author = {{Ramirez}, Ramses M.},
        title = "{The effect of high nitrogen pressures on the habitable zone and an appraisal of greenhouse states}",
      journal = {\mnras},
         year = 2020,
        month = may,
       volume = {494},
       number = {1},
        pages = {259-270},
          doi = {10.1093/mnras/staa603},
archivePrefix = {arXiv},
       eprint = {2004.00229},
 primaryClass = {astro-ph.EP},
       adsurl = {https://ui.adsabs.harvard.edu/abs/2020MNRAS.494..259R}
}

@ARTICLE{2019ApJ...881..120K,
       author = {{Koll}, Daniel D.~B. and {Cronin}, Timothy W.},
        title = "{Hot Hydrogen Climates Near the Inner Edge of the Habitable Zone}",
      journal = {\apj},
         year = 2019,
        month = aug,
       volume = {881},
       number = {2},
          eid = {120},
        pages = {120},
          doi = {10.3847/1538-4357/ab30c4},
archivePrefix = {arXiv},
       eprint = {1907.13169},
 primaryClass = {astro-ph.EP},
       adsurl = {https://ui.adsabs.harvard.edu/abs/2019ApJ...881..120K}
}

@ARTICLE{2022A&A...658A..40C,
       author = {{Chaverot}, Guillaume and {Turbet}, Martin and {Bolmont}, Emeline and {Leconte}, J{\'e}r{\'e}my},
        title = "{How does the background atmosphere affect the onset of the runaway greenhouse?}",
      journal = {\aap},
         year = 2022,
        month = feb,
       volume = {658},
          eid = {A40},
        pages = {A40},
          doi = {10.1051/0004-6361/202142286},
archivePrefix = {arXiv},
       eprint = {2111.07662},
 primaryClass = {astro-ph.EP},
       adsurl = {https://ui.adsabs.harvard.edu/abs/2022A&A...658A..40C}
}

@ARTICLE{1983AmJS..283..641B,
       author = {{Berner}, R.~A. and {Lasaga}, A.~C. and {Garrels}, R.~M.},
        title = "{The carbonate-silicate geochemical cycle and its effect on atmospheric carbon dioxide over the past 100 million years}",
      journal = {American Journal of Science},
         year = 1983,
        month = sep,
       volume = {283},
       number = {7},
        pages = {641-683},
          doi = {10.2475/ajs.283.7.641},
       adsurl = {https://ui.adsabs.harvard.edu/abs/1983AmJS..283..641B}
}

@ARTICLE{1981JGR....86.9776W,
       author = {{Walker}, J.~C.~G. and {Hays}, P.~B. and {Kasting}, J.~F.},
        title = "{A negative feedback mechanism for the long-term stabilization of the earth's surface temperature}",
      journal = {\jgr},
         year = 1981,
        month = oct,
       volume = {86},
        pages = {9776-9782},
          doi = {10.1029/JC086iC10p09776},
       adsurl = {https://ui.adsabs.harvard.edu/abs/1981JGR....86.9776W}
}

@ARTICLE{2016A&A...592A..36G,
       author = {{Godolt}, M. and {Grenfell}, J.~L. and {Kitzmann}, D. and {Kunze}, M. and {Langematz}, U. and {Patzer}, A.~B.~C. and {Rauer}, H. and {Stracke}, B.},
        title = "{Assessing the habitability of planets with Earth-like atmospheres with 1D and 3D climate modeling}",
      journal = {\aap},
         year = 2016,
        month = jul,
       volume = {592},
          eid = {A36},
        pages = {A36},
          doi = {10.1051/0004-6361/201628413},
archivePrefix = {arXiv},
       eprint = {1605.08231},
 primaryClass = {astro-ph.EP},
       adsurl = {https://ui.adsabs.harvard.edu/abs/2016A&A...592A..36G}
}

@ARTICLE{2013ApJ...765..131K,
       author = {{Kopparapu}, Ravi Kumar and {Ramirez}, Ramses and {Kasting}, James F. and {Eymet}, Vincent and {Robinson}, Tyler D. and {Mahadevan}, Suvrath and {Terrien}, Ryan C. and {Domagal-Goldman}, Shawn and {Meadows}, Victoria and {Deshpande}, Rohit},
        title = "{Habitable Zones around Main-sequence Stars: New Estimates}",
      journal = {\apj},
         year = 2013,
        month = mar,
       volume = {765},
       number = {2},
          eid = {131},
        pages = {131},
          doi = {10.1088/0004-637X/765/2/131},
archivePrefix = {arXiv},
       eprint = {1301.6674},
 primaryClass = {astro-ph.EP},
       adsurl = {https://ui.adsabs.harvard.edu/abs/2013ApJ...765..131K}
}

@ARTICLE{2025PSJ.....6..212L,
       author = {{Landry}, Jared and {Kurokawa}, Hiroyuki and {Taki}, Tetsuo and {Fujii}, Yuka and {Aoki}, Kosuke and {Genda}, Hidenori},
        title = "{Impacts of Atmospheric Carbon Species and Stellar Type on Climates of Terrestrial Planets}",
      journal = {\psj},
         year = 2025,
        month = sep,
       volume = {6},
       number = {9},
          eid = {212},
        pages = {212},
          doi = {10.3847/PSJ/adf7a1},
archivePrefix = {arXiv},
       eprint = {2508.05975},
 primaryClass = {astro-ph.EP},
       adsurl = {https://ui.adsabs.harvard.edu/abs/2025PSJ.....6..212L}
}

@ARTICLE{Arney+2016AsBio..16..873A,
       author = {{Arney}, Giada and {Domagal-Goldman}, Shawn D. and {Meadows}, Victoria S. and {Wolf}, Eric T. and {Schwieterman}, Edward and {Charnay}, Benjamin and {Claire}, Mark and {H{\'e}brard}, Eric and {Trainer}, Melissa G.},
        title = "{The Pale Orange Dot: The Spectrum and Habitability of Hazy Archean Earth}",
      journal = {Astrobiology},
         year = 2016,
        month = nov,
       volume = {16},
       number = {11},
        pages = {873-899},
          doi = {10.1089/ast.2015.1422},
archivePrefix = {arXiv},
       eprint = {1610.04515},
 primaryClass = {astro-ph.EP},
       adsurl = {https://ui.adsabs.harvard.edu/abs/2016AsBio..16..873A}
}

@ARTICLE{Keles+2018AsBio..18..116K,
       author = {{Keles}, Engin and {Grenfell}, John Lee and {Godolt}, Mareike and {Stracke}, Barbara and {Rauer}, Heike},
        title = "{The Effect of Varying Atmospheric Pressure upon Habitability and Biosignatures of Earth-like Planets}",
      journal = {Astrobiology},
         year = 2018,
        month = feb,
       volume = {18},
       number = {2},
        pages = {116-132},
          doi = {10.1089/ast.2016.1632},
archivePrefix = {arXiv},
       eprint = {2006.05207},
 primaryClass = {astro-ph.EP},
       adsurl = {https://ui.adsabs.harvard.edu/abs/2018AsBio..18..116K}
}

@ARTICLE{Seeley+Wordsworth2023PSJ.....4...34S,
       author = {{Seeley}, Jacob T. and {Wordsworth}, Robin D.},
        title = "{Moist Convection Is Most Vigorous at Intermediate Atmospheric Humidity}",
      journal = {\psj},
         year = 2023,
        month = feb,
       volume = {4},
       number = {2},
          eid = {34},
        pages = {34},
          doi = {10.3847/PSJ/acb0cb},
archivePrefix = {arXiv},
       eprint = {2301.03669},
 primaryClass = {astro-ph.EP},
       adsurl = {https://ui.adsabs.harvard.edu/abs/2023PSJ.....4...34S}
}

@ARTICLE{Watanabe+Ozaki2024ApJ...961....1W,
       author = {{Watanabe}, Yasuto and {Ozaki}, Kazumi},
        title = "{Relative Abundances of CO$_{2}$, CO, and CH$_{4}$ in Atmospheres of Earth-like Lifeless Planets}",
      journal = {\apj},
         year = 2024,
        month = jan,
       volume = {961},
       number = {1},
          eid = {1},
        pages = {1},
          doi = {10.3847/1538-4357/ad10a2},
archivePrefix = {arXiv},
       eprint = {2309.13538},
 primaryClass = {astro-ph.EP},
       adsurl = {https://ui.adsabs.harvard.edu/abs/2024ApJ...961....1W}
}

@ARTICLE{vonParis2013P&SS...82..149V,
       author = {{von Paris}, P. and {Grenfell}, J.~L. and {Rauer}, H. and {Stock}, J.~W.},
        title = "{N$_{2}$-associated surface warming on early Mars}",
      journal = {\planss},
         year = 2013,
        month = jul,
       volume = {82},
        pages = {149-154},
          doi = {10.1016/j.pss.2013.04.009},
archivePrefix = {arXiv},
       eprint = {1304.6024},
 primaryClass = {astro-ph.EP},
       adsurl = {https://ui.adsabs.harvard.edu/abs/2013P&SS...82..149V}
}

@ARTICLE{1989JGR....9416287T,
       author = {{Toon}, Owen B. and {McKay}, C.~P. and {Ackerman}, T.~P. and {Santhanam}, K.},
        title = "{Rapid calculation of radiative heating rates and photodissociation rates in inhomogeneous multiple scattering atmospheres}",
      journal = {\jgr},
         year = 1989,
        month = nov,
       volume = {94},
        pages = {16287-16301},
          doi = {10.1029/JD094iD13p16287},
       adsurl = {https://ui.adsabs.harvard.edu/abs/1989JGR....9416287T}
}

@ARTICLE{1999JQSRT..62..109K,
       author = {{Kato}, S. and {Ackerman}, T.~P. and {Mather}, J.~H. and {Clothiaux}, E.~E.},
        title = "{The k-distribution method and correlated-k approximation for a shortwave radiative transfer model.}",
      journal = {\jqsrt},
         year = 1999,
        month = may,
       volume = {62},
       number = {1},
        pages = {109-121},
          doi = {10.1016/S0022-4073(98)00075-2},
       adsurl = {https://ui.adsabs.harvard.edu/abs/1999JQSRT..62..109K}
}

@ARTICLE{Kurokawa+2014E&PSL.394..179K,
       author = {{Kurokawa}, H. and {Sato}, M. and {Ushioda}, M. and {Matsuyama}, T. and {Moriwaki}, R. and {Dohm}, J.~M. and {Usui}, T.},
        title = "{Evolution of water reservoirs on Mars: Constraints from hydrogen isotopes in martian meteorites}",
      journal = {Earth and Planetary Science Letters},
         year = 2014,
        month = may,
       volume = {394},
        pages = {179-185},
          doi = {10.1016/j.epsl.2014.03.027},
archivePrefix = {arXiv},
       eprint = {1403.4211},
 primaryClass = {astro-ph.EP},
       adsurl = {https://ui.adsabs.harvard.edu/abs/2014E&PSL.394..179K}
}

@ARTICLE{Goldblatt+2013NatGe...6..661G,
       author = {{Goldblatt}, Colin and {Robinson}, Tyler D. and {Zahnle}, Kevin J. and {Crisp}, David},
        title = "{Low simulated radiation limit for runaway greenhouse climates}",
      journal = {Nature Geoscience},
         year = 2013,
        month = aug,
       volume = {6},
       number = {8},
        pages = {661-667},
          doi = {10.1038/ngeo1892},
       adsurl = {https://ui.adsabs.harvard.edu/abs/2013NatGe...6..661G}
}

@ARTICLE{Abraham+Goldblatt2022JAtS...79.2243A,
       author = {{Abraham}, Carsten and {Goldblatt}, Colin},
        title = "{A Satellite Climatology of Relative Humidity Profiles and Outgoing Thermal Radiation over Earth's Oceans}",
      journal = {Journal of the Atmospheric Sciences},
         year = 2022,
        month = sep,
       volume = {79},
       number = {9},
        pages = {2243-2265},
          doi = {10.1175/JAS-D-21-0270.1},
       adsurl = {https://ui.adsabs.harvard.edu/abs/2022JAtS...79.2243A}
}

@ARTICLE{Kite+2021PNAS..11801959K,
       author = {{Kite}, Edwin S. and {Steele}, Liam J. and {Mischna}, Michael A. and {Richardson}, Mark I.},
        title = "{Warm early Mars surface enabled by high-altitude water ice clouds}",
      journal = {Proceedings of the National Academy of Science},
         year = 2021,
        month = may,
       volume = {118},
       number = {18},
          eid = {e2101959118},
        pages = {e2101959118},
          doi = {10.1073/pnas.2101959118},
       adsurl = {https://ui.adsabs.harvard.edu/abs/2021PNAS..11801959K}
}

@ARTICLE{Ueno+2024NatGe..17..503U,
       author = {{Ueno}, Yuichiro and {Schmidt}, Johan A. and {Johnson}, Matthew S. and {Zang}, Xiaofeng and {Gilbert}, Alexis and {Kurokawa}, Hiroyuki and {Usui}, Tomohiro and {Aoki}, Shohei},
        title = "{Synthesis of $^{13}$C-depleted organic matter from CO in a reducing early Martian atmosphere}",
      journal = {Nature Geoscience},
         year = 2024,
        month = jun,
       volume = {17},
       number = {6},
        pages = {503-507},
          doi = {10.1038/s41561-024-01443-z},
       adsurl = {https://ui.adsabs.harvard.edu/abs/2024NatGe..17..503U}
}

@ARTICLE{Ramirez+Craddock2018NatGe..11..230R,
       author = {{Ramirez}, Ramses M. and {Craddock}, Robert A.},
        title = "{The geological and climatological case for a warmer and wetter early Mars}",
      journal = {Nature Geoscience},
         year = 2018,
        month = apr,
       volume = {11},
       number = {4},
        pages = {230-237},
          doi = {10.1038/s41561-018-0093-9},
archivePrefix = {arXiv},
       eprint = {1810.01974},
 primaryClass = {astro-ph.EP},
       adsurl = {https://ui.adsabs.harvard.edu/abs/2018NatGe..11..230R}
}

@ARTICLE{Ehlmann+2016JGRE..121.1927E,
       author = {{Ehlmann}, B.~L. and {Anderson}, F.~S. and {Andrews-Hanna}, J. and {Catling}, D.~C. and {Christensen}, P.~R. and {Cohen}, B.~A. and {Dressing}, C.~D. and {Edwards}, C.~S. and {Elkins-Tanton}, L.~T. and {Farley}, K.~A. and {Fassett}, C.~I. and {Fischer}, W.~W. and {Fraeman}, A.~A. and {Golombek}, M.~P. and {Hamilton}, V.~E. and {Hayes}, A.~G. and {Herd}, C.~D.~K. and {Horgan}, B. and {Hu}, R. and {Jakosky}, B.~M. and {Johnson}, J.~R. and {Kasting}, J.~F. and {Kerber}, L. and {Kinch}, K.~M. and {Kite}, E.~S. and {Knutson}, H.~A. and {Lunine}, J.~I. and {Mahaffy}, P.~R. and {Mangold}, N. and {McCubbin}, F.~M. and {Mustard}, J.~F. and {Niles}, P.~B. and {Quantin-Nataf}, C. and {Rice}, M.~S. and {Stack}, K.~M. and {Stevenson}, D.~J. and {Stewart}, S.~T. and {Toplis}, M.~J. and {Usui}, T. and {Weiss}, B.~P. and {Werner}, S.~C. and {Wordsworth}, R.~D. and {Wray}, J.~J. and {Yingst}, R.~A. and {Yung}, Y.~L. and {Zahnle}, K.~J.},
        title = "{The sustainability of habitability on terrestrial planets: Insights, questions, and needed measurements from Mars for understanding the evolution of Earth-like worlds}",
      journal = {Journal of Geophysical Research (Planets)},
         year = 2016,
        month = oct,
       volume = {121},
       number = {10},
        pages = {1927-1961},
          doi = {10.1002/2016JE005134},
       adsurl = {https://ui.adsabs.harvard.edu/abs/2016JGRE..121.1927E}
}

@ARTICLE{Kurokawa+2018Icar..299..443K,
       author = {{Kurokawa}, Hiroyuki and {Kurosawa}, Kosuke and {Usui}, Tomohiro},
        title = "{A lower limit of atmospheric pressure on early Mars inferred from nitrogen and argon isotopic compositions}",
      journal = {\icarus},
         year = 2018,
        month = jan,
       volume = {299},
        pages = {443-459},
          doi = {10.1016/j.icarus.2017.08.020},
archivePrefix = {arXiv},
       eprint = {1708.03956},
 primaryClass = {astro-ph.EP},
       adsurl = {https://ui.adsabs.harvard.edu/abs/2018Icar..299..443K}
}

@ARTICLE{Jakosky+1994Icar..111..271J,
       author = {{Jakosky}, Bruce M. and {Pepin}, Robert O. and {Johnson}, Robert E. and {Fox}, J.~L.},
        title = "{Mars Atmospheric Loss and Isotopic Fractionation by Solar-Wind-Induced Sputtering and Photochemical Escape}",
      journal = {\icarus},
         year = 1994,
        month = oct,
       volume = {111},
       number = {2},
        pages = {271-288},
          doi = {10.1006/icar.1994.1145},
       adsurl = {https://ui.adsabs.harvard.edu/abs/1994Icar..111..271J}
}

@ARTICLE{Turbet+2021,
       author = {{Turbet}, Martin and {Bolmont}, Emeline and {Chaverot}, Guillaume and {Ehrenreich}, David and {Leconte}, J{\'e}r{\'e}my and {Marcq}, Emmanuel},
        title = "{Day-night cloud asymmetry prevents early oceans on Venus but not on Earth}",
      journal = {\nat},
         year = 2021,
        month = oct,
       volume = {598},
       number = {7880},
        pages = {276-280},
          doi = {10.1038/s41586-021-03873-w},
archivePrefix = {arXiv},
       eprint = {2110.08801},
 primaryClass = {astro-ph.EP},
       adsurl = {https://ui.adsabs.harvard.edu/abs/2021Natur.598..276T}
}

@ARTICLE{Hamano+2013,
       author = {{Hamano}, Keiko and {Abe}, Yutaka and {Genda}, Hidenori},
        title = "{Emergence of two types of terrestrial planet on solidification of magma ocean}",
      journal = {\nat},
         year = 2013,
        month = may,
       volume = {497},
       number = {7451},
        pages = {607-610},
          doi = {10.1038/nature12163},
       adsurl = {https://ui.adsabs.harvard.edu/abs/2013Natur.497..607H}
}

@ARTICLE{Gillman+2020,
       author = {{Gillmann}, C. and {Golabek}, G.~J. and {Raymond}, S.~N. and {Sch{\"o}nb{\"a}chler}, M. and {Tackley}, P.~J. and {Dehant}, V. and {Debaille}, V.},
        title = "{Dry late accretion inferred from Venus's coupled atmosphere and internal evolution}",
      journal = {Nature Geoscience},
         year = 2020,
        month = apr,
       volume = {13},
       number = {4},
        pages = {265-269},
          doi = {10.1038/s41561-020-0561-x},
archivePrefix = {arXiv},
       eprint = {2010.07132},
 primaryClass = {astro-ph.EP},
       adsurl = {https://ui.adsabs.harvard.edu/abs/2020NatGe..13..265G}
}

@ARTICLE{Gillmann+2009,
       author = {{Gillmann}, C{\'e}dric and {Chassefi{\`e}re}, Eric and {Lognonn{\'e}}, Philippe},
        title = "{A consistent picture of early hydrodynamic escape of Venus atmosphere explaining present Ne and Ar isotopic ratios and low oxygen atmospheric content}",
      journal = {Earth and Planetary Science Letters},
         year = 2009,
        month = sep,
       volume = {286},
       number = {3-4},
        pages = {503-513},
          doi = {10.1016/j.epsl.2009.07.016},
       adsurl = {https://ui.adsabs.harvard.edu/abs/2009E&PSL.286..503G}
}

@ARTICLE{Kasting+Ackerman1986Sci...234.1383K,
       author = {{Kasting}, James F. and {Ackerman}, Thomas P.},
        title = "{Climatic Consequences of Very High Carbon Dioxide Levels in the Earth's Early Atmosphere}",
      journal = {Science},
         year = 1986,
        month = dec,
       volume = {234},
       number = {4782},
        pages = {1383-1385},
          doi = {10.1126/science.11539665},
       adsurl = {https://ui.adsabs.harvard.edu/abs/1986Sci...234.1383K}
}

@ARTICLE{Ramirez+2014AsBio..14..714R,
       author = {{Ramirez}, Ramses M. and {Kopparapu}, Ravi Kumar and {Lindner}, Valerie and {Kasting}, James F.},
        title = "{Can Increased Atmospheric CO2 Levels Trigger a Runaway Greenhouse?}",
      journal = {Astrobiology},
         year = 2014,
        month = aug,
       volume = {14},
       number = {8},
        pages = {714-731},
          doi = {10.1089/ast.2014.1153},
       adsurl = {https://ui.adsabs.harvard.edu/abs/2014AsBio..14..714R}
}

@ARTICLE{Manabe+Wetherald1967,
       author = {{Manabe}, Syukuro and {Wetherald}, Richard T.},
        title = "{Thermal Equilibrium of the Atmosphere with a Given Distribution of Relative Humidity.}",
      journal = {Journal of the Atmospheric Sciences},
         year = 1967,
        month = may,
       volume = {24},
       number = {3},
        pages = {241-259},
          doi = {10.1175/1520-0469(1967)024<0241:TEOTAW>2.0.CO;2},
       adsurl = {https://ui.adsabs.harvard.edu/abs/1967JAtS...24..241M}
}

@ARTICLE{Sholes+2017Icar..290...46S,
       author = {{Sholes}, Steven F. and {Smith}, Megan L. and {Claire}, Mark W. and {Zahnle}, Kevin J. and {Catling}, David C.},
        title = "{Anoxic atmospheres on Mars driven by volcanism: Implications for past environments and life}",
      journal = {\icarus},
         year = 2017,
        month = jul,
       volume = {290},
        pages = {46-62},
          doi = {10.1016/j.icarus.2017.02.022},
archivePrefix = {arXiv},
       eprint = {2103.13012},
 primaryClass = {astro-ph.EP},
       adsurl = {https://ui.adsabs.harvard.edu/abs/2017Icar..290...46S}
}

@ARTICLE{zahnle2008photochemical,
       author = {{Zahnle}, Kevin and {Haberle}, Robert M. and {Catling}, David C. and {Kasting}, James F.},
        title = "{Photochemical instability of the ancient Martian atmosphere}",
      journal = {Journal of Geophysical Research (Planets)},
         year = 2008,
        month = nov,
       volume = {113},
       number = {E11},
          eid = {E11004},
        pages = {E11004},
          doi = {10.1029/2008JE003160},
       adsurl = {https://ui.adsabs.harvard.edu/abs/2008JGRE..11311004Z}
}

@ARTICLE{Hu+2020,
       author = {{Hu}, Renyu and {Peterson}, Luke and {Wolf}, Eric T.},
        title = "{O$_{2}$- and CO-rich Atmospheres for Potentially Habitable Environments on TRAPPIST-1 Planets}",
      journal = {\apj},
         year = 2020,
        month = jan,
       volume = {888},
       number = {2},
          eid = {122},
        pages = {122},
          doi = {10.3847/1538-4357/ab5f07},
archivePrefix = {arXiv},
       eprint = {1912.02313},
 primaryClass = {astro-ph.EP},
       adsurl = {https://ui.adsabs.harvard.edu/abs/2020ApJ...888..122H}
}

@ARTICLE{harman2015abiotic,
       author = {{Harman}, C.~E. and {Schwieterman}, E.~W. and {Schottelkotte}, J.~C. and {Kasting}, J.~F.},
        title = "{Abiotic O$_{2}$ Levels on Planets around F, G, K, and M Stars: Possible False Positives for Life?}",
      journal = {\apj},
         year = 2015,
        month = oct,
       volume = {812},
       number = {2},
          eid = {137},
        pages = {137},
          doi = {10.1088/0004-637X/812/2/137},
archivePrefix = {arXiv},
       eprint = {1509.07863},
 primaryClass = {astro-ph.EP},
       adsurl = {https://ui.adsabs.harvard.edu/abs/2015ApJ...812..137H}
}

@ARTICLE{Tian+2014,
       author = {{Tian}, Feng and {France}, Kevin and {Linsky}, Jeffrey L. and {Mauas}, Pablo J.~D. and {Vieytes}, Mariela C.},
        title = "{High stellar FUV/NUV ratio and oxygen contents in the atmospheres of potentially habitable planets}",
      journal = {Earth and Planetary Science Letters},
         year = 2014,
        month = jan,
       volume = {385},
        pages = {22-27},
          doi = {10.1016/j.epsl.2013.10.024},
archivePrefix = {arXiv},
       eprint = {1310.2590},
 primaryClass = {astro-ph.EP},
       adsurl = {https://ui.adsabs.harvard.edu/abs/2014E&PSL.385...22T}
}

@ARTICLE{Luger+Barnes2015AsBio..15..119L,
       author = {{Luger}, R. and {Barnes}, R.},
        title = "{Extreme Water Loss and Abiotic O2Buildup on Planets Throughout the Habitable Zones of M Dwarfs}",
      journal = {Astrobiology},
         year = 2015,
        month = feb,
       volume = {15},
       number = {2},
        pages = {119-143},
          doi = {10.1089/ast.2014.1231},
archivePrefix = {arXiv},
       eprint = {1411.7412},
 primaryClass = {astro-ph.EP},
       adsurl = {https://ui.adsabs.harvard.edu/abs/2015AsBio..15..119L}
}

@ARTICLE{Lammer+2019AsBio..19..927L,
       author = {{Lammer}, Helmut and {Spro{\ss}}, Laurenz and {Grenfell}, John Lee and {Scherf}, Manuel and {Fossati}, Luca and {Lendl}, Monika and {Cubillos}, Patricio E.},
        title = "{The Role of N$_{2}$ as a Geo-Biosignature for the Detection and Characterization of Earth-like Habitats}",
      journal = {Astrobiology},
         year = 2019,
        month = jul,
       volume = {19},
       number = {7},
        pages = {927-950},
          doi = {10.1089/ast.2018.1914},
archivePrefix = {arXiv},
       eprint = {1904.11716},
 primaryClass = {astro-ph.EP},
       adsurl = {https://ui.adsabs.harvard.edu/abs/2019AsBio..19..927L}
}

@ARTICLE{Way+DelGenio2020JGRE..12506276W,
       author = {{Way}, M.~J. and {Del Genio}, Anthony D.},
        title = "{Venusian Habitable Climate Scenarios: Modeling Venus Through Time and Applications to Slowly Rotating Venus-Like Exoplanets}",
      journal = {Journal of Geophysical Research (Planets)},
         year = 2020,
        month = may,
       volume = {125},
       number = {5},
          eid = {e06276},
        pages = {e06276},
          doi = {10.1029/2019JE00627610.1002/essoar.10501118.3},
       adsurl = {https://ui.adsabs.harvard.edu/abs/2020JGRE..12506276W}
}

@ARTICLE{Way+2016GeoRL..43.8376W,
       author = {{Way}, M.~J. and {Del Genio}, Anthony D. and {Kiang}, Nancy Y. and {Sohl}, Linda E. and {Grinspoon}, David H. and {Aleinov}, Igor and {Kelley}, Maxwell and {Clune}, Thomas},
        title = "{Was Venus the first habitable world of our solar system?}",
      journal = {\grl},
         year = 2016,
        month = aug,
       volume = {43},
       number = {16},
        pages = {8376-8383},
          doi = {10.1002/2016GL069790},
archivePrefix = {arXiv},
       eprint = {1608.00706},
 primaryClass = {astro-ph.EP},
       adsurl = {https://ui.adsabs.harvard.edu/abs/2016GeoRL..43.8376W}
}

@article{adams2025episodic,
  title={Episodic warm climates on early Mars primed by crustal hydration},
  author={Adams, Danica and Scheucher, Markus and Hu, Renyu and Ehlmann, Bethany L and Thomas, Trent B and Wordsworth, Robin and Scheller, Eva and Lillis, Rob and Smith, Kayla and Rauer, Heike and others},
  journal={Nature Geoscience},
  pages={1--7},
  year={2025},
  publisher={Nature Publishing Group UK London}
}

@ARTICLE{Thomas+2023PSJ.....4...41T,
       author = {{Thomas}, Trent B. and {Hu}, Renyu and {Lo}, Daniel Y.},
        title = "{Constraints on the Size and Composition of the Ancient Martian Atmosphere from Coupled CO$_{2}$-N$_{2}$-Ar Isotopic Evolution Models}",
      journal = {\psj},
         year = 2023,
        month = mar,
       volume = {4},
       number = {3},
          eid = {41},
        pages = {41},
          doi = {10.3847/PSJ/acb924},
archivePrefix = {arXiv},
       eprint = {2302.04241},
 primaryClass = {astro-ph.EP},
       adsurl = {https://ui.adsabs.harvard.edu/abs/2023PSJ.....4...41T}
}

@ARTICLE{Hu+Thomas2022NatGe..15..106H,
       author = {{Hu}, Renyu and {Thomas}, Trent B.},
        title = "{A nitrogen-rich atmosphere on ancient Mars consistent with isotopic evolution models}",
      journal = {Nature Geoscience},
         year = 2022,
        month = feb,
       volume = {15},
       number = {2},
        pages = {106-111},
          doi = {10.1038/s41561-021-00886-y},
archivePrefix = {arXiv},
       eprint = {2202.04825},
 primaryClass = {astro-ph.EP},
       adsurl = {https://ui.adsabs.harvard.edu/abs/2022NatGe..15..106H}
}

@ARTICLE{Vladilo+2013ApJ...767...65V,
       author = {{Vladilo}, Giovanni and {Murante}, Giuseppe and {Silva}, Laura and {Provenzale}, Antonello and {Ferri}, Gaia and {Ragazzini}, Gregorio},
        title = "{The Habitable Zone of Earth-like Planets with Different Levels of Atmospheric Pressure}",
      journal = {\apj},
         year = 2013,
        month = apr,
       volume = {767},
       number = {1},
          eid = {65},
        pages = {65},
          doi = {10.1088/0004-637X/767/1/65},
archivePrefix = {arXiv},
       eprint = {1302.4566},
 primaryClass = {astro-ph.EP},
       adsurl = {https://ui.adsabs.harvard.edu/abs/2013ApJ...767...65V}
}

@ARTICLE{Wordsworth+Pierrehumbert2014ApJ...785L..20W,
       author = {{Wordsworth}, Robin and {Pierrehumbert}, Raymond},
        title = "{Abiotic Oxygen-dominated Atmospheres on Terrestrial Habitable Zone Planets}",
      journal = {\apjl},
         year = 2014,
        month = apr,
       volume = {785},
       number = {2},
          eid = {L20},
        pages = {L20},
          doi = {10.1088/2041-8205/785/2/L20},
archivePrefix = {arXiv},
       eprint = {1403.2713},
 primaryClass = {astro-ph.EP},
       adsurl = {https://ui.adsabs.harvard.edu/abs/2014ApJ...785L..20W}
}

@ARTICLE{Kurokawa+2022GGG....2310295K,
       author = {{Kurokawa}, H. and {Laneuville}, M. and {Li}, Y. and {Zhang}, N. and {Fujii}, Y. and {Sakuraba}, H. and {Houser}, C. and {Cleaves}, H.~J.},
        title = "{The Origin of Earth's Mantle Nitrogen: Primordial or Early Biogeochemical Cycling?}",
      journal = {Geochemistry, Geophysics, Geosystems},
         year = 2022,
        month = may,
       volume = {23},
       number = {5},
          eid = {e2021GC010295},
        pages = {e2021GC010295},
          doi = {10.1029/2021GC010295},
archivePrefix = {arXiv},
       eprint = {2204.14002},
 primaryClass = {physics.geo-ph},
       adsurl = {https://ui.adsabs.harvard.edu/abs/2022GGG....2310295K}
}

@ARTICLE{Sakuraba+2021NatSR..1120894S,
       author = {{Sakuraba}, Haruka and {Kurokawa}, Hiroyuki and {Genda}, Hidenori and {Ohta}, Kenji},
        title = "{Numerous chondritic impactors and oxidized magma ocean set Earth's volatile depletion}",
      journal = {Scientific Reports},
         year = 2021,
        month = oct,
       volume = {11},
          eid = {20894},
        pages = {20894},
          doi = {10.1038/s41598-021-99240-w},
archivePrefix = {arXiv},
       eprint = {2110.12195},
 primaryClass = {astro-ph.EP},
       adsurl = {https://ui.adsabs.harvard.edu/abs/2021NatSR..1120894S}
}

@article{hirschmann2016constraints,
  title={Constraints on the early delivery and fractionation of Earth's major volatiles from C/H, C/N, and C/S ratios},
  author={Hirschmann, Marc M},
  journal={American Mineralogist},
  volume={101},
  number={3},
  pages={540--553},
  year={2016}
}

@ARTICLE{Johnson+Goldblatt2015ESRv..148..150J,
       author = {{Johnson}, Ben and {Goldblatt}, Colin},
        title = "{The nitrogen budget of Earth}",
      journal = {Earth Science Reviews},
         year = 2015,
        month = sep,
       volume = {148},
        pages = {150-173},
          doi = {10.1016/j.earscirev.2015.05.006},
archivePrefix = {arXiv},
       eprint = {1505.03813},
 primaryClass = {astro-ph.EP},
       adsurl = {https://ui.adsabs.harvard.edu/abs/2015ESRv..148..150J}
}

@ARTICLE{Marty2012E&PSL.313...56M,
       author = {{Marty}, Bernard},
        title = "{The origins and concentrations of water, carbon, nitrogen and noble gases on Earth}",
      journal = {Earth and Planetary Science Letters},
         year = 2012,
        month = jan,
       volume = {313},
        pages = {56-66},
          doi = {10.1016/j.epsl.2011.10.040},
archivePrefix = {arXiv},
       eprint = {1405.6336},
 primaryClass = {astro-ph.EP},
       adsurl = {https://ui.adsabs.harvard.edu/abs/2012E&PSL.313...56M}
}

@ARTICLE{Marty+2013Sci...342..101M,
       author = {{Marty}, Bernard and {Zimmermann}, Laurent and {Pujol}, Magali and {Burgess}, Ray and {Philippot}, Pascal},
        title = "{Nitrogen Isotopic Composition and Density of the Archean Atmosphere}",
      journal = {Science},
         year = 2013,
        month = oct,
       volume = {342},
       number = {6154},
        pages = {101-104},
          doi = {10.1126/science.1240971},
archivePrefix = {arXiv},
       eprint = {1405.6337},
 primaryClass = {astro-ph.EP},
       adsurl = {https://ui.adsabs.harvard.edu/abs/2013Sci...342..101M}
}

@ARTICLE{Kavanagh+Goldblatt2015E&PSL.413...51K,
       author = {{Kavanagh}, Lucas and {Goldblatt}, Colin},
        title = "{Using raindrops to constrain past atmospheric density}",
      journal = {Earth and Planetary Science Letters},
         year = 2015,
        month = mar,
       volume = {413},
        pages = {51-58},
          doi = {10.1016/j.epsl.2014.12.032},
       adsurl = {https://ui.adsabs.harvard.edu/abs/2015E&PSL.413...51K}
}

@ARTICLE{Som+2016NatGe...9..448S,
       author = {{Som}, Sanjoy M. and {Buick}, Roger and {Hagadorn}, James W. and {Blake}, Tim S. and {Perreault}, John M. and {Harnmeijer}, Jelte P. and {Catling}, David C.},
        title = "{Earth's air pressure 2.7 billion years ago constrained to less than half of modern levels}",
      journal = {Nature Geoscience},
         year = 2016,
        month = jun,
       volume = {9},
       number = {6},
        pages = {448-451},
          doi = {10.1038/ngeo2713},
       adsurl = {https://ui.adsabs.harvard.edu/abs/2016NatGe...9..448S}
}

@ARTICLE{Som+2012Natur.484..359S,
       author = {{Som}, Sanjoy M. and {Catling}, David C. and {Harnmeijer}, Jelte P. and {Polivka}, Peter M. and {Buick}, Roger},
        title = "{Air density 2.7 billion years ago limited to less than twice modern levels by fossil raindrop imprints}",
      journal = {\nat},
         year = 2012,
        month = apr,
       volume = {484},
       number = {7394},
        pages = {359-362},
          doi = {10.1038/nature10890},
       adsurl = {https://ui.adsabs.harvard.edu/abs/2012Natur.484..359S}
}

@BOOK{catling2017atmospheric,
       author = {{Catling}, David C. and {Kasting}, James F.},
        title = "{Atmospheric Evolution on Inhabited and Lifeless Worlds}",
         year = 2017,
       adsurl = {https://ui.adsabs.harvard.edu/abs/2017aeil.book.....C}
}

@ARTICLE{2013ApJ...778..154W,
       author = {{Wordsworth}, R.~D. and {Pierrehumbert}, R.~T.},
        title = "{Water Loss from Terrestrial Planets with CO$_{2}$-rich Atmospheres}",
      journal = {\apj},
         year = 2013,
        month = dec,
       volume = {778},
       number = {2},
          eid = {154},
        pages = {154},
          doi = {10.1088/0004-637X/778/2/154},
archivePrefix = {arXiv},
       eprint = {1306.3266},
 primaryClass = {astro-ph.EP},
       adsurl = {https://ui.adsabs.harvard.edu/abs/2013ApJ...778..154W}
}

@ARTICLE{2009NatGe...2..891G,
       author = {{Goldblatt}, Colin and {Claire}, Mark W. and {Lenton}, Timothy M. and {Matthews}, Adrian J. and {Watson}, Andrew J. and {Zahnle}, Kevin J.},
        title = "{Nitrogen-enhanced greenhouse warming on early Earth}",
      journal = {Nature Geoscience},
         year = 2009,
        month = dec,
       volume = {2},
       number = {12},
        pages = {891-896},
          doi = {10.1038/ngeo692},
       adsurl = {https://ui.adsabs.harvard.edu/abs/2009NatGe...2..891G}
}

@ARTICLE{1988Icar...74..472K,
       author = {{Kasting}, J.~F.},
        title = "{Runaway and moist greenhouse atmospheres and the evolution of Earth and Venus}",
      journal = {\icarus},
         year = 1988,
        month = jun,
       volume = {74},
       number = {3},
        pages = {472-494},
          doi = {10.1016/0019-1035(88)90116-9},
       adsurl = {https://ui.adsabs.harvard.edu/abs/1988Icar...74..472K}
}
\bibliographystyle{aasjournal}

\end{document}